
\input phyzzx
\hfuzz=35pt
\FRONTPAGE
\line{\hfill BROWN-HET-893}
\line{\hfill February 1993}
\vskip1.0truein
\titlestyle{{ LECTURES ON MODERN COSMOLOGY AND STRUCTURE FORMATION
}\foot{Invited lectures at the 7th Swieca Summer School in Particles
and Fields, Campos do Jord\~ao, Brazil, January 10-23, 1993; to be
published in the proceedings, eds. O. Eboli and V. Ribelles (World
Scientific, Singapore, 1993).}}
\bigskip
\author{Robert H. BRANDENBERGER}
\centerline{{\it Department of Physics}}
\centerline{{\it Brown University, Providence, RI 02912, USA}}
\bigskip
\abstract
Focus of these lectures is the challenge of explaining the origin of
structure in the Universe.  The interplay between quantum field theory
and classical general relativity has given rise to several interesting
cosmological models which contain mechanisms for generating density
inhomogeneities.  The three theories discussed here are the
inflationary Universe, the cosmic string and the global texture
models.  The recent COBE discovery of anisotropies in the microwave
background has provided some support for all three models.  The
present results do not allow a distinction between these models.
Statistics which distinguish between the predictions of the three
theories are discussed.
\endpage

\chapter{Introduction}
\par
The marriage of quantum field theory and general relativity has over
the past decade led to significant progress in cosmology.  For the
first time, theories have been developed which provide causal
explanations for the origin and distribution of structure in the
Universe.

In these lectures I discuss inflationary Universe and topological
defect models of cosmology, focusing on the question of structure
formation.  These notes are intended as a pedagogical introduction
rather than as a comprehensive review.  For comprehensive discussions
of inflation the reader is referred to Refs. 1-3, and for detailed
reviews to topological defect models to Refs. 4-7.  An introduction to
quantum field theory methods used in modern cosmology can be found in
Ref. 8.  These notes draw on material presented elsewhere$^{9,10)}$ in
which some of the topics are treated in more detail.

When speaking of structure in the Universe, I will concentrate on
cosmic structures of galactic scales and larger, scales on which
complicated nonlinear and hydrodynamical effects can be neglected.  We
would like to understand the reasons for the particular nonrandom
distribution of galaxies and galaxy clusters manifested, e.g., in
the ``Great Wall" of galaxies discovered in the Center for
Astrophysics redshift survey$^{11)}$.  The theory must also be able to
explain the recently discovered anisotropies in the cosmic microwave
background (CMB) temperature$^{12)}$.

The ``standard big bang" cosmology, in spite of its success in
predicting the existence and spectral form of the CMB and in
explaining the abundance of light elements in the Universe, leaves all
questions concerning inhomogeneities in the Universe unanswered.  This
``old" cosmology was based on classical general relativity as the
theory of space and time, and on hydrodynamical matter.  In about 1980
it was realized that a ``new" cosmology based on quantum field matter
but maintaining a classical description of space-time in terms of
general relativity leads to theories which generate density
fluctuations and hence provide possible explanations for the observed
distribution of matter.

It is rather natural to adopt the ``new" cosmology.  After all, since
we know that ``old" cosmology predicts that the Universe started out
very small and hot, the ideal gas description of matter will
inevitably break down at sufficiently early times.  At that point, we
should adopt the theory which describes matter best at high energies
and temperature, namely quantum field theory.

However, a fundamental limitation of these ``new" cosmological models
is already apparent: treating half of the system (space-time)
classically and only the other half (matter) quantum mechanically is
intrinsically inconsistent.  This is why, in spite of their successes
at explaining many observations, the present theories should only be
regarded as toy models or effective theories which are desperately
seeking a unified basis.  A second major deficiency of the present
models is that they do not address the cosmological constant problem.
The theories I will discuss all involve scalar matter fields whose
potential energy must be carefully adjusted (fine tuned) to vanish at the
global
minimum.

The outline of these lectures is as follows.  Section 2 is a review of
standard cosmology.  I focus on its conceptual bases and observational
support, and close with a discussion of a few of its major problems
and of their resolution in the context of inflation$^{13)}$.

Section 3 is an overview of the inflationary Universe scenario.  I
explain why it is possible to obtain inflation if matter is described
in terms of scalar fields.  After a brief field theory interlude
covering finite temperature effects and phase transitions I review the
early mechanisms to obtain inflation and the reasons why they fail.
The chapter closes with a discussion of the most simple self
consistent model -- chaotic inflation$^{14)}$.

One of the most important achievements of inflation is to provide a
mechanism for the generation of density inhomogeneities.  As is
explained in Section 4, this mechanisms is based on quantum vacuum
fluctuations of the inflaton field which generate scalar-type metric
perturbations which are most efficiently studied using the gauge
invariant theory of cosmological perturbations$^{15)}$.

A quite different mechanism of structure formation is based on
topological defects which arise during phase transitions in the very
early Universe$^{16)}$.  In Section 5 I give a classification of
topological defects, I then compare cosmic strings and global
textures, the latter an example of a ``semi-topological defect."

In Section 6, the topological defect models of structure formation are
developed.  Similarities and differences between the predictions of
the inflationary Universe scenario, the cosmic string models and the
texture theory are explained.  The recent COBE detections$^{12)}$ of
anisotropies in the CMB only give information about quantities which
do not distinguish between the three models.  I will discuss recent
attempts to find statistics which allow a clean distinction.

I devoted my final lecture at the Swieca School to present a new way
of resolving one of the outstanding problems of modern cosmology, the
singularity problem.  Since this lecture was beyond the main focus of
the lecture series, I do not include it here, but refer the reader to
the original article$^{17)}$ and to a separate review$^{18)}$.

In this writeup, units in which $c = \hbar = k_B = 1$ are used unless
mentioned otherwise.  The metric $g_{\mu\nu}$ is taken to have
signature $(+, -, -, -)$.  Greek indices run over space and time,
latin ones over spatial indices only.  The Hubble expansion rate is
$H(t) = \dot a (t) / a(t)$, with $a(t)$ the scale factor of a
Friedmann-Robertson-Walker (FRW) Universe.  The present value of $H$
is $h 100$ kms$^{-1}$ Mpc$^{-1}$, where $0.4 < h < 1$.  Unless stated
otherwise, the value of $h$ is taken to be 0.5.
With $z(t)$ the cosmological redshift at time $t$ is denoted. As usual, the
symbols
$G$ and $m_{pl}$ stand for Newton's constant and Planck mass,
respectively.  Distances are measured in pc or Mpc, where 1 pc
corresponds to 3.1 light years.

\chapter{Standard Cosmology:  Successes and Problems}

The standard big bang cosmology rests on three theoretical pillars:
the cosmological principle, Einstein's general theory of relativity
and a perfect fluid description of matter.

The cosmological principle$^{19)}$ states that on large distance
scales the Universe is homogeneous.  From an observational point of view
this is an extremely nontrivial statement.  On small scales the
Universe looks extremely inhomogeneous.  The inhomogeneities of the
solar system are obvious to everyone, and even by naked eye it is
apparent that stars are not randomly distributed.  They are bound into
galaxies, dynamical entities whose visible radius is about 10$^4$ pc.
Telescopic observations show that galaxies are not randomly
distributed, either.  Dense clumps of galaxies can be identified as
Abell clusters.  In turn, Abell cluster positions are correlated to
produce the large-scale structure dominated by sheets (or filaments)
with typical scale 100 Mpc observed in recent redshift
surveys$^{11)}$.  Until recently, every new survey probing the
Universe to greater depth revealed new structures on the scale of the
sample volume.  In terms of the visible distribution of matter there
was no evidence for large-scale homogeneity.  This situation changed
in 1992 with the announcement$^{20)}$ that a new redshift survey
complete to a depth of about 800 Mpc had discovered no prominent
structures on scales larger than 100 Mpc.  This is the first
observational evidence from optical measurements in favor of the
cosmological principle.  However, to put this result in perspective we
must keep in mind that the observed isotropy of the CMB temperature to
better than 10$^{-5}$ on large angular scales has been excellent
evidence for the validity of the cosmological principle.

The second theoretical pillar is general relativity, the theory which
determines the dynamics of the Universe.  According to general
relativity, space-time is a smooth manifold.  Together with the
cosmological principle this tells us that it is possible to choose a
family of hypersurfaces with maximal symmetry.  These are the
homogeneous constant time hypersurfaces.  The metric of these surfaces
is $^{21)}$
$$
ds^2 = a(t)^2 \, \left[ {dr^2\over{1-kr^2}} + r^2 (d\theta^2 + \sin^2
\theta \, d \varphi^2) \right] \eqno\eq
$$
when using spherical polar coordinates.  The constant $k$ is $+1$, $0$
or $-1$ for closed, flat or open surfaces respectively.  The function
$a(t)$ is the scale factor of the Universe.  By a coordinate choice,
it could be set equal to 1 at any given time.  However, the time
dependence of $a(t)$ indicates how the spatial sections evolve as a
function of time.  The full space-time metric is
$$
ds^2 = dt^2 - a (t)^2 \, \left[ {dr^2\over{1-kr^2}} + r^2 (d\theta^2 + \sin^2
\theta \, d \varphi^2) \right] \, . \eqno\eq
$$
According to the Einstein equivalence principle, test particles move
on geodesics with respect to the metric given by $ds^2$.  This implies
that the peculiar velocity $\underline{v}_p$ obeys the equation
$$
\underline{\dot v}_p + {\dot a\over a} \, \underline{v}_p = 0 \, . \eqno\eq
$$
Here,
$$
\underline{v}_p \equiv a (t) \, {d \underline{x}^c\over{dt}} \, , \eqno\eq
$$
$\underline{x}^c$ being the comoving coordinates $r$, $\theta$ and $\varphi$.
Equation (2.3) implies that in an expanding Universe, the peculiar
velocity $\underline{v}_p$ decreases:
$$
\underline{v}_p (t) \sim a^{-1} (t) \, . \eqno\eq
$$
Therefore, trajectories with constant $\underline{x}^c$ are geodesics and
correspond to particles at rest.  The velocity $\underline{v}_p$ is the
physical
velocity relative to the expansion of the Universe (see Fig. 1).
\goodbreak \midinsert \vskip 4.5cm
\hsize=6in \raggedright
\noindent {\bf Figure 1.} Sketch of the expanding Universe.
Concentric circles indicate space at fixed time, time increasing as
the radius gets larger.  Points at rest have constant comoving
coordinates.  Their world lines are straight lines through the origin
(e.g. $L$).
\endinsert

 From observations$^{22)}$ it is known that the Universe is at present
expanding.  Looking at distant galaxies, we detect a redshift $z$ of
light which increases linearly with the distance $d$ of the
object:
$$
z (d) \sim d \, , \eqno\eq
$$
where the redshift $z$ is defined by
$$
z = {\lambda_0\over \lambda_e} - 1 \, , \eqno\eq
$$
with $\lambda_0$ and $\lambda_e$ being the wavelengths measured by us
and by the source.  The relationship (2.6) is explained by taking
galaxies to be at rest in comoving coordinates, and $a(t)$ to be
increasing.  In this case for $z \ll 1$
$$
z \simeq H (t_0) d \>\> \left[ H (t) = {\dot a \over a} (t) \right] \,
, \eqno\eq
$$
$t_0$ being the present time.

The most important consequence of general relativity for the history
of the Universe is that it relates the expansion rate to the matter
content.  The Einstein field equations follow from the action
$$
S = \int d^4 x \, \sqrt{-g} \, (R - 16 \pi G {\cal L}_M) \eqno\eq
$$
where $R$ is the Ricci scalar curvature, $g$ is the determinant of the
metric, and ${\cal L}_M$ is the Lagrange density for matter.
Evaluating the equations of motion obtained by varying (2.9) with
respect to $g_{\mu \nu}$ for a metric of the form (2.2) leads to the
famous Friedmann-Robertson-Walker (FRW) equations
$$
\left( {\dot a\over a} \right)^2 - {k\over a^2} = {8 \pi G\over 3} \,
\rho \eqno\eq
$$
$$
{\ddot a\over a} = - {4 \pi G\over 3} \> (\rho + 3p) \, . \eqno\eq
$$
Note that for fixed energy density $\rho$, the evolution of $a(t)$
depends in an important way on the pressure $p$.

Besides explaining Hubble's redshift-distance relation (2.6), standard
big bang cosmology makes two key quantitative predictions:  the
existence of a black body cosmic microwave background$^{23)}$ and
nucleosynthesis, the generation of light elements$^{23-25)}$.

Consider ordinary matter made up of atoms in an expanding Universe.
The energy density in matter scales as $a(t)^{-3}$, and the
temperature $T (t)$ as $a^{-1}(t)$.  Thus, standard cosmology predicts
that as we go back in time, the Universe was warmer.  In particular,
at a critical temperature $T_{rec}$, matter becomes ionized.  Before
$t_{rec}$ (the time corresponding to $T_{rec}$) the Universe was
opaque to photons, after $t_{rec}$ it was transparent.  To be more
precise, $T_{rec}$ is the temperature when photons fall out of thermal
equilibrium.  Thereafter they propagate without scattering.  The black
body nature of the spectrum of photons is maintained, but the
temperature redshifts.  Hence, the big bang model predicts$^{23,26)}$
a black body spectrum of photons of temperature
$$
T_0 = T_{rec} \, z^{-1}_{rec} \eqno\eq
$$
where the cosmological redshift $z(t)$ is given by
$$
1 + z (t) = {a (t_0)\over{a (t)}} \eqno\eq
$$
and $z_{rec} = z (t_{rec}) \sim 10^3$.

In 1965, Penzias and Wilson$^{27)}$ discovered this remnant black body
radiation at a temperature of about $3^\circ$ K.  Since the spectrum
peaks in the microwave region it is now called CMB (cosmic microwave
background).  Recent satellite (COBE)$^{28)}$ and rocket $^{29)}$
experiments have confirmed the black body nature of the CMB to very
high accuracy.  The temperature is 2.73$^\circ$ K $=T_0$.

Given the existence of the CMB, we know that matter has two
components: dust (with energy density $\rho_m (t)$) and radiation
(with density $\rho_r (t)$).  At the present time $t_0$, $\rho_m (t)
\gg \rho_r (t)$.  The radiation energy density is determined by $T_0$,
and the matter energy density can be estimated by analyzing the
dynamics of galaxies and clusters and using the virial theorem.  However, since
$$
\eqalign{
& \rho_m (t) \sim a (t)^{-3} \cr
& \rho_r (t) \sim a (t)^{-4} \, , } \eqno\eq
$$
as we go back in time the fraction of energy density in radiation
increases, and the two components become equal at $t_{eq}$, the time of
equal matter and radiation.  The corresponding redshift is
$$
z_{eq} = \Omega h^{-2}_{50} \, 10^4 \eqno\eq
$$
where
$$
\Omega = {\rho\over{\rho_{cr}}} (t_0) \, , \eqno\eq
$$
$\rho_{cr}$ being the density for a spatially flat Universe (the
critical density), and $h_{50}$ in the value of $H$ in units of 50 km
s$^{-1}$ Mpc$^{-1}$.

The time $t_{eq}$ is important for structure formation$^{30)}$.  It is
only after $t_{eq}$ that perturbations on scales smaller than the
Hubble radius $H^{-1} (t)$ can grow.  Before then, the radiation
pressure prevents growth.  A temperature-time plot of the early
Universe is sketched in Fig. 2.  Note that $t_{eq} < t_{rec}$.
\goodbreak \midinsert \vskip 6.5cm
\hsize=6in \raggedright
\noindent {\bf Figure 2.} Temperature-time diagram of standard big
bang cosmology.  The present time, time of decoupling and time of
equal matter and radiation are $t_0$, $t_{rec}$ and $t_{eq}$,
respectively.  The Universe is radiation dominated before $t_{eq}$
(Region A), and matter dominated in Region B.  Before and after
$t_{rec}$ the Universe is opaque and transparent, respectively, to
microwave photons.
\endinsert

The second quantitative prediction of standard big bang cosmology
concerns nucleosynthesis.  Above a temperature of about $10^{9 \circ}$
K, the nuclear interactions are sufficiently fast to prevent neutrons
and protons from fusing.  However, below that temperature, it is
thermodynamically favorable$^{23-25)}$ for neutrons and protons to
fuse and form deuterium, helium 3, helium 4 and lithium 7 through a
long and interconnected chain of reactions.  The resulting light
element abundances depend sensitively on the expansion rate of the
Universe and on $\Omega_B$, the fraction of energy density $\rho_B$ at
present in baryons relative to the critical density $\rho_c$.  In Fig.
3, recent$^{31)}$ theoretical calculations of the abundances are shown
and compared with observations.  Demanding agreement with all
abundances leaves only a narrow window
\goodbreak \midinsert \vskip 12.5cm
\hsize=6in \raggedright
\noindent {\bf Figure 3.} Light element abundances as a function of
the baryon to entropy ratio $\eta$.  $Y_p$ is the $^4$He mass
fraction, calculated assuming three light neutrino species, for two
different values $\tau_{1/2}$ of the neutrino half life (in minutes).
The abundances of $D + ^3$He and of $^7$Li are shown as ratios of
their number density relative to the number density of $H$.  The
horizontal lines indicate limits from observations: upper limit of
10$^{-4}$ for the $D + ^3$He abundance, and upper limits on $^7$Li
from observations of dwarf stars.  The $^7$Li curve is shown with $\pm
2 \sigma$ errors.  Combining $^4$He and $^3$He $+ D$ limits leaves
only a small window for $\eta$ which is allowed.  A major success of
primordial nucleosynthesis is that the $^7$Li abundance matches well.
Note that $\eta_{10}$ is $\eta$ in units of 10$^{-10}$.
\endinsert
$$
3 \times 10^{-10} < \eta < 10^{-9} \, , \eqno\eq
$$
where $\eta$ is the ratio of baryon number density $n_B$ to entropy
density $s$
$$
\eta = {n_B\over s} \, . \eqno\eq
$$
 From (2.17), it follows that $\Omega_B$ is constrained:
$$
0.01 < \Omega_B h^2 < 0.035 \, . \eqno\eq
$$
In particular, if the Universe is spatially flat, there must be
nonbaryonic dark matter.  We will return to the dark matter issue
shortly.

In summary, the three observational pillars of standard big bang
cosmology are Hubble's redshift-distance relation, the existence and
black body spectrum of the CMB, and the concordance between observed
and theoretically determined light element abundances.

Standard big bang cosmology is faced with several important problems:
the age, dark matter, homogeneity, flatness, and formation of
structure problems (this list is not exhaustive$^{32)}$, but contains
what I consider to be the key problems).  In addition, standard big
bang cosmology does not explain the small value of the cosmological
constant -- a problem which no present cosmological model addresses
(see, however, Ref. 33 for some recent progress on this issue).  Of
the above five problems, only the first is a possible conflict of the
theory with observations.  The other four are questions which are left
unanswered by the theory.  Extensions of the standard model are needed
to address these issues.

Globular cluster ages have been estimated to lie in the range $13 - 18
\times 10^9$ years$^{34)}$.  Nuclear cosmochronology gives an age of
structures in the range $(10 - 20) \times 10^9$ yr.  However, big bang
cosmology (in the absence of a cosmological constant) predicts an age
$$
\tau = {7\over h} \cdot 10^2 {\rm yr} \, . \eqno\eq
$$
Thus, theory and observations are only compatible if $h < 0.55$.

The dark matter problem$^{35)}$ problem has various aspects.  There is
more matter in galaxies than is visible in stars.  This follows by
studying galactic velocity rotation curves and observing that the
velocity remains constant beyond the visible radius of the galaxy.
Whereas$^{34)}$ the contribution of stars to $\Omega$ is less than
$0.01$, galaxies as a whole contribute $ > 0.02$ to $\Omega$.  The
second level of the dark matter problem lies in clusters.  By studying
cluster dynamics it can be inferred that the contribution to $\Omega$
exceeds $0.1$.  Finally, observations on the largest scales give an
even larger contribution to $\Omega$.  From Virgo infall it follows
that $\Omega > 0.3$, and from large scale velocity measurements
(POTENT)$^{36)}$ or from infrared galaxy (IRAS) surveys$^{37)}$ it
follows that
$$
0.5 < \Omega < 3 \, . \eqno\eq
$$
As we shall see in the next section, inflation predicts $\Omega \simeq
1$, unless fine tuned initial conditions are chosen$^{38)}$.

The bottom line of the dark matter problem is that more mass is
observed than can be in light (hence the name ``dark matter"), and
that most of this missing matter must be nonbaryonic (since by (2.19)
$\Omega_B < 0.14$).  Note that there must be some baryonic dark
matter, and that all of the galactic dark matter could be baryonic.
This comment will be relevant when discussing cosmic strings and
galaxy formation in Section 6.  If $h = 0.5$, then most of the cluster
dark matter must be nonbaryonic.  Standard cosmology does not address
the issue of what the dark matter is.  New cosmology can provide a
solution.  Quantum field theory models of matter in the Universe
give rise to several candidate particles which could constitute the dark
matter$^{39)}$.

The last three problems are the classic problems which the
inflationary Universe scenario addresses$^{13)}$.  In Fig. 4, the
homogeneity (or horizon) problem is illustrated.  As is sketched, the
region $\ell_p (t_{eq})$ over which the CMB is observed to be homogeneous
to better than one part in $10^5$ is much larger than the forward
light cone $\ell_f (t_{eq})$ at $t_{eq}$, which is the maximal
distance over which microphysical forces could have caused the
homogeneity:
$$
\ell_p (t_{eq}) = a(t_{eq}) \, \int\limits^{t_0}_{t_{eq}} \, dt \, a^{-1} (t)
\simeq 3 t_0 \left( 1 - \left({t_{eq}\over t_0} \right)^{1/2} \right)
\eqno\eq
$$
$$
\ell_f (t_{eq}) = a (t_{eq}) \int\limits^{t_{eq}}_0 \, dt \, a^{-1}
(t) \simeq 3 t_0^{2/3} \, t^{1/3}_{eq} \, . \eqno\eq
$$
 From the above equations it is obvious that $\ell_p (t_{eq}) \gg
\ell_f (t_{eq})$.  Hence, standard cosmology cannot explain the
observed isotropy of the CMB.
\goodbreak \midinsert \vskip 7.5cm
\hsize=6in \raggedright
\noindent {\bf Figure 4.} A sketch of the homogeneity problem: the
past light cone $\ell_p (t)$ at the time $t_{rec}$ of last scattering
is much larger than the forward light cone $\ell_f (t)$ at $t_{rec}$.
\endinsert

In standard cosmology and in an expanding Universe, $\Omega = 1$ is an
unstable fixed point.  This can be seen as follows.  For a spatially
flat Universe $(\Omega = 1)$
$$
H^2 = {8 \pi G\over 3} \, \rho_c \, , \eqno\eq
$$
whereas for a nonflat Universe
$$
H^2 + \varepsilon \, T^2 = \, {8 \pi G\over 3} \, \rho \, , \eqno\eq
$$
with
$$
\varepsilon = {k \over{(a T)^2}} \, . \eqno\eq
$$
The quantity $\varepsilon$ is proportional to $s^{-2/3}$, where $s$ is
the entropy density.  Hence, in standard cosmology, $\varepsilon$ is
constant.  Combining (2.24) and (2.25) gives
$$
{\rho - \rho_c\over \rho_c} = {3 \over{8 \pi G}} \> {\varepsilon T^2
\over \rho_c} \sim T^{-2} \, . \eqno\eq
$$
Thus, as the temperature decreases, $\Omega - 1$ increases.  In fact,
in order to explain the present small value of $\Omega - 1 \sim {\cal O}
(1)$, the initial energy density had to be extremely close to critical
density.  For example, at $T= 10^{15}$ GeV, (2.27) implies
$$
{\rho - \rho_c\over \rho_c} \sim 10^{-50} \, . \eqno\eq
$$
What is the origin of these fine tuned initial conditions?  This is
the flatness problem of standard cosmology.
\par
The last problem of the standard cosmological model I will mention is the
``formation of structure problem."  Observations indicate that galaxies and
even clusters of galaxies have nonrandom correlations on scales
larger than 50 Mpc$^{11,40,41)}$.  This scale is comparable to the comoving
horizon at $t_{eq}$. Thus,
if the initial density perturbations were produced much before $t_{eq}$, the
correlations cannot be explained by a causal mechanism.  Gravity alone is, in
general, too weak to build up correlations on the scale of clusters after
$t_{eq}$ (see, however, the explosion scenario of Ref. 42).  Hence, the two
questions of what generates the
primordial density perturbations and what causes the observed correlations, do
not have an answer in the context of standard cosmology.  This problem is
illustrated by Fig. 5.
\goodbreak\midinsert \vskip 7cm
\hsize=6in \raggedright
\noindent {\bf Figure 5.} A sketch (conformal separation vs. time) of the
formation of structure problem:  the comoving separation $d_c$ between two
clusters is larger than the forward light cone at time $t_{\rm rec}$.
\endinsert
\par
In 1981, based on previous work by many other people (see e.g., Refs.
1-3 for a
detailed bibliography of early work), Guth realized$^{13)}$ that having a
sufficiently long phase in the very early Universe during which the scale
factor expands exponentially
$$
a (t) \sim e^{Ht}\eqno\eq
$$
can potentially solve the three last problems listed above.  This phase of
exponential expansion is called the de Sitter or inflationary phase.
\topinsert \vskip 11cm
\hsize=6in \raggedright
\noindent {\bf Figure 6.} A sketch (physical coordinates vs. time) of the
solution of the homogeneity problem.  Due to exponential expansion the forward
light cone is larger than the past light cone at $t_{\rm rec}$.  The dashed
line is the apparent horizon or Hubble radius.
\endinsert
\par
In Fig. 6 it is sketched how a period of inflation can solve the homogeneity
problem.  $t_i$ shall denote the onset of inflation, $t_R$ the end.
$\triangle t = t_R - t_i$ is the period of inflation.  During inflation, the
forward light cone increases exponentially compared to a model without
inflation, whereas the past light cone is not affected for $t \ge t_R$.
Hence, provided $\triangle t$ is sufficiently large, $\ell_f (t_{eq})$ will be
greater than $\ell_p^c (t_R)$.  The condition on $\triangle t$ depends on the
temperature $T_R$ corresponding to time $t_R$, the temperature of reheating.
Demanding that $\ell_f (t_R) > \ell_p (t_R)$ we find, using (2.22) and
(2.23), the following criterion
$$
e^{\triangle t H} \ge \, {\ell_p(t_R)\over{\ell_f (t_R)}} \, \simeq \,
\left({t_0\over{t_R}}\right)^{1/2} \, = \, \left({T_R\over{T_0}}\right) \,
\sim \, 10^{27} \eqno\eq
$$
for $T_R \sim 10^{14}$GeV and $T_0 \sim 10^{-13}$GeV (the present microwave
background temperature).  Thus, in order to solve the homogeneity problem, a
period of inflation with
$$
\triangle t > 50 \, H^{-1}\eqno\eq
$$
is required.
\par
Inflation also can solve the flatness problem$^{13,43)}$.  The key point is
that the
entropy density $s$ is no longer constant.  As will be explained later, the
temperatures at $t_i$ and $t_R$ are essentially equal.  Hence, the entropy
increases during inflation by a factor $\exp (3 H \triangle t)$.  Thus,
$\epsilon$ decreases by a factor of $\exp (-2 H \triangle t)$.  With the
numbers used in (2.30):
$$
\epsilon_{\rm after} \sim 10^{-54} \epsilon_{\rm before}\, . \eqno\eq
$$
Hence, $(\rho - \rho_c)/\rho$ can be of order 1 both at $t_i$ and at the
present time.  In fact,if inflation occurs at all, the theory then predicts
that
at the present time $\Omega = 1$ to a high accuracy (now $\Omega < 1$ would
require special initial conditions).
\par
What was said above can be rephrased geometrically:  during inflation, the
curvature radius of the Universe -- measured on a fixed physical scale --
increases exponentially.  Thus, a piece of space looks essentially flat after
inflation even if it had measurable curvature before.
\par
Most importantly, inflation provides a mechanism which in a causal way
generates the primordial perturbations required for galaxies, clusters and
even larger objects$^{44-47)}$.  In inflationary Universe models, the
``apparent" horizon
$3t$ and the ``actual" horizon (the forward light cone) do not coincide at
late times.  Provided (2.30) is
satisfied, then (as sketched in Fig. 7) all scales within our apparent horizon
were inside the actual
horizon since $t_i$.  Thus, it is in principle possible to have a causal
generation mechanism for perturbations.
\goodbreak \midinsert \vskip 10.5cm
\hsize=6in \raggedright
\noindent {\bf Figure 7.} A sketch (physical coordinates vs. time) of the
solution of the formation of structure problem.  The separation $d_c$ between
two clusters is always smaller than the forward light cone.  The dashed line
is the Hubble radius $H^{-1} (t)$.
\endinsert
\par
\chapter{The Inflationary Universe}
\section{Inflation and Generalized Inflation}
\par
An inflationary phase is a period in time during which $a(t)$ increases
exponentially.  For many purposes, however, a less restrictive definition is
adequate.  It is sufficient to require that $H^{-1} (t)$ decreases in comoving
coordinates.  As illustrated in Fig. 8, this is sufficient to ensure that
scales of galaxies and clusters originate inside the Hubble radius at some
early time.
\midinsert \vskip 6.5cm
\hsize=6in \raggedright
\noindent {\bf Figure 8.}  The Hubble radius $H^{-1} (t)$ as a function of
time in a model of generalized inflation.  Also plotted is a fixed comoving
scale $d_c$.
\endinsert
\par
Assume a power law growth of the scale factor,
$$
a (t) \sim t^q\eqno\eq
$$
Then, in comoving coordinates, the Hubble radius is
$$
H^{-1}_c (t) \sim t^{1 -q}\eqno\eq
$$
Thus, for generalized inflation, $q > 1$, i.e., $\ddot a > 0$, is required.
This requirement can be translated into a condition on the equation of state.
 From (2.11) it follows that $\ddot a > 0$ only if
$$
p < - {1\over 3} \rho\, .\eqno\eq
$$
In particular, negative pressure is required in order to obtain generalized
inflation.  Thus, we must go beyond the ideal gas description of matter to
obtain inflation.  This is one of the reasons for turning to a description of
matter in terms of classical and quantum fields.
\par
There is another reason for turning to a description of matter in terms of
quantum fields.  The standard cosmological model predicts that as we go
backwards in time the temperature increases.  It is now well established that
at high temperatures (or equivalently energies) the classical description of
matter as an ideal gas breaks down.  Hence, we expect that an improved
description of the early Universe can be obtained by using the theory of
matter which holds at very high energies.  Currently, the prime candidate is
quantum field theory.
\par
Using quantum (or even classical) fields, it is in principle possible to obtain
inflation.  To verify this, consider a scalar field $\varphi (\undertext{x},
t)$.
The associated energy density $\rho (\undertext{x}, t)$ and pressure
$p(\undertext{x},t)$ can be determined from the classical energy-momentum
tensor $T_{\mu\nu} (\varphi (\undertext{x}, t))$:  In an expanding Universe
with
metric
$$
g_{\mu\nu} \, = \,  {\rm diag} \, \left(1, - a^2 (t), -a^2 (t), -a^2
(t)\right)\eqno\eq
$$
and for a Lagrangian
$$
{\cal L} \, = \, {1\over 2} \partial_\mu \varphi \partial^\mu \varphi -
V (\varphi)\eqno\eq
$$
with potential energy density $V(\varphi)$ we obtain
$$
\eqalign{\rho (\undertext{x}, t) & = \, {1\over 2} \dot \varphi^2 \,
(\undertext{x}, t) + \, {1\over 2} a^{-2} \, (\bigtriangledown \varphi)^2 +
V(\varphi)\cr
p (\undertext{x}, t) & = \, {1\over 3} \sum^3_{i = 1} T_{ii} \, = \, {1\over
2} \dot \varphi^2 \, (\undertext{x}, t) - {1\over 6} \, a^{-2}
(\bigtriangledown \varphi)^2 - V (\varphi)\, .}\eqno\eq
$$
Thus, if $\varphi (\undertext{x}, t_i) =$const
and $\dot \varphi (\undertext{x}, t_i) = 0$ at some initial time $t_i$
and $V (\varphi (\undertext{x}, t_i)) > 0$,
then the equation of state becomes $p = - \rho$ and leads to inflation.
\par
\midinsert \vskip 5cm
\hsize=6in \raggedright
\noindent {\bf Figure 9.}  A sketch of two potentials which can give rise to
inflation.
\endinsert
Two examples which give inflation are shown in Fig. 9.  In (a), inflation
occurs at the stable fixed point
$\varphi (\undertext{x}, t_i) = 0 = \dot \varphi (\undertext{x}, t_i)$.
However, this model is ruled out by observation: the inflationary
phase has no ending.  $V(0)$ acts as a permanent nonvanishing cosmological
constant.  In (b), a finite period of inflation can arise if
$\varphi(\undertext{x})$ is trapped at the local minimum $\varphi = 0$ with
$\dot \varphi (\undertext{x}) = 0$.  However, in this case $\varphi
(\undertext{x})$ can make a sudden transition at some time $t_R > t_i$ through
the potential barrier and move to $\varphi (\undertext{x}) = a$.  Thus, for
$t_i < t < t_R$ the Universe expands exponentially, whereas for $t > t_R$ the
contribution of $\varphi$ to the expansion of the Universe vanishes and we get
the usual FRW cosmology.  There are two obvious questions: how does the
transition occur and why should the scalar field have $V(\varphi) = 0$ at the
global minimum.  In the following section the first question will be
addressed.  The second question is part of the cosmological constant problem
for which there is as yet no convincing explanation.  Before studying
the dynamics of the phase transition, we need to digress and discuss
finite temperature effects.
\section{Finite Temperature Field Theory}
\par
The evolution of particles in vacuum and in a thermal bath are very different.
Similarly, the evolution of fields changes when coupled to a thermal bath.
Under certain conditions, the changes may be absorbed in a temperature
dependent potential, the finite temperature effective
potential$^{48)}$.
Here, a heuristic derivation of this potential will be given. The reader is
referred to Ref. 8 or to the original articles$^{48)}$ for the actual
derivation.

        We assume that the scalar field $\varphi (\undertext{x},t)$ is
coupled to a thermal bath which is represented by a second scalar field
$\psi (\undertext{x}, t)$ in thermal equilibrium. The
Lagrangian for $\varphi$ is
$$
{\cal L} = {1\over 2} \partial_\mu \varphi \partial^\mu \varphi \, - \,
V(\varphi) \, - \, {1\over 2} \hat \lambda \varphi^2 \psi^2\, , \eqno\eq
$$
where $\hat \lambda$ is a coupling constant.  The action from which the
equations of motion are derived is
$$
S \, = \, \int d^4 x \, \sqrt{-g} {\cal L}\eqno\eq
$$
where $g$ is the determinant of the metric (3.4).   The resulting equation of
motion for $\varphi (\undertext{x},t)$ is
$$
\ddot \varphi + 3 H \dot \varphi \, - a^{-2} \bigtriangledown^2 \varphi \, =
\,- V^\prime (\varphi) - \hat \lambda \psi^2 \varphi \, . \eqno\eq
$$
If $\psi$ is in thermal equilibrium, we may replace $\psi^2$ by its thermal
expectation value $<\psi^2>_T$.  Now,
$$
<\psi^2>_T \sim T^2\eqno\eq
$$
which can be seen as follows:  in thermal equilibrium, the energy density of
$\psi$ equals that of one degree of freedom in the thermal bath.  In
particular, the potential energy density $V (\psi)$ of $\psi$ is of that order
of magnitude.  Let
$$
V (\psi) \, = \, \lambda_\psi \psi^4\eqno\eq
$$
with a coupling constant $\lambda_\psi$ which we take to be of the order 1
(if $\lambda_\psi$ is too small, $\psi$ will not be in thermal equilibrium).
Since the thermal energy density is proportional to $T^4$, (3.10) follows.
(3.9) can be rewritten as
$$
\ddot \psi + 3H \dot \varphi \, - \, a^{-2} \bigtriangledown^2 \varphi \, = \,
- V_T^\prime (\varphi),\eqno\eq
$$
where
$$
V_T (\varphi) \, = \, V (\varphi) \, + \, {1\over 2} \hat \lambda T^2
\varphi^2\eqno\eq
$$
is called the finite temperature effective potential.  Note that in
(3.13),
$\hat \lambda$ has been rescaled to absorb the constant of proportionality in
(3.10).
\par
These considerations will now be applied to Example A, a scalar field model
with
potential
$$
V (\varphi) \, = \, {1\over 4} \lambda (\varphi^2 - \sigma^2)^2\eqno\eq
$$
($\sigma$ is called the scale of symmetry breaking).  The finite temperature
effective potential becomes (see Fig. 10)
$$
V_T (\varphi) \, = \, {1\over 4} \lambda \varphi^4 - {1\over 2}
\, \left(\lambda \sigma^2 - \hat \lambda T^2\right) \varphi^2 +
\, {1\over 4} \, \lambda \sigma^4 \, . \eqno\eq
$$
For very high temperatures, the effective mass term is positive
and hence the energetically favorable state is $<\varphi> = 0$.
For very low temperatures, on the other hand, the mass term has a
negative sign which leads to spontaneous symmetry breaking.  The
temperature at which the mass term vanishes defines the critical
temperature $T_c$
$$
T_c \, = \, \hat \lambda^{-1/2} \lambda^{1/2} \sigma\,.\eqno\eq
$$
\midinsert \vskip 6cm
\hsize=6in \raggedright
\noindent {\bf Figure 10.}  The finite temperature effective potential
for Example A.
\endinsert
\hfuzz=15pt
\par
As Example B, consider a theory with potential
$$
 V (\varphi) = \, {1\over 4} \varphi^4 - {1\over 3} \, (a + b) \varphi^3 +
{1\over 2} ab \varphi^2\eqno\eq
$$
with ${1\over 2} a > b > 0$.  The finite temperature effective potential is
obtained by adding ${1\over 2} \hat \lambda  T^2 \varphi^2$ to the
rhs of (3.17).  ${ V_T} (\varphi)$
is sketched in Fig. 11 for various values of $ T$.  The critical temperature
$T_c$ is defined as the temperature when the two minima of ${V_T}(\varphi)$
become degenerate.
\par
\midinsert \vskip 6cm
\hsize=6in \raggedright
\noindent {\bf Figure 11.} The finite temperature effective potential
for Example B.
\endinsert
It is important to note that the use of finite temperature effective potential
methods is only legitimate if the system is in thermal equilibrium.  This
point was stressed in Refs. 14 and 49, although the conclusion should be
obvious from the derivation given above.  To be more precise, we require the
$\psi$ field to be in thermal equilibrium and the coupling constant $\hat
\lambda$ of (3.7) which mediates the energy exchange between the $\varphi$
and $\psi$ fields to be large.  However, as shown in Chapter 4, observational
constraints stemming from the amplitude of the primordial energy density
fluctuation spectrum force the self coupling constant $\lambda$ of $\varphi$
to be extremely small.  Since at one loop order, the interaction term ${1\over
2} \hat \lambda \varphi^2 \psi^2$ induces contributions to $\lambda$, it is
unnatural to have $\lambda$ very small and $\hat \lambda$ unsuppressed.
Hence, in many inflationary Universe models-in particular in new
inflation$^{50)}$ and in chaotic inflation$^{14)}$ - finite temperature
effective potential methods are inapplicable.
\section{Phase Transitions}
\par
The temperature dependence of the finite temperature effective potential in
quantum field theory leads to phase transitions in the very early Universe.
These transitions are either first or second order.
\par
Example $A$ of the previous section provides a model in which the transition
is second order (see Fig. 10).  For $T \gg T_c$, the expectation value of the
scalar field $\varphi$ vanishes at all points $\undertext{x}$ in space:
$$
< \varphi (\undertext{x}) > = 0 \, . \eqno\eq
$$
For $T < T_c$, this value of $< \varphi (\undertext{x}) >$ becomes unstable
and $< \varphi (\undertext{x})>$ evolves smoothly in time to a new value $\pm
\sigma$.  The direction is determined by thermal and quantum fluctuations and
is therefore not uniform in space.  There will be domains of average radius
$\xi (t)$ in which $< \varphi (\undertext{x}) >$ is coherent.  By causality,
the coherence length is bounded from above by the horizon.  However, typical
values of $\xi (t)$ are proportional to $\lambda^{-1} \sigma^{-1}$ if
$\varphi$ was in thermal equilibrium before the phase
transition$^{16)}$.
\par
In condensed matter physics, a transition of the above type is said to proceed
by spinodal decomposition$^{51)}$, triggered by a rapid quench.
\par
In Example B of the previous section, (see Fig. 11) the phase
transition is first order.  For $T > T_c$, the expectation value
$<\varphi (x) >$ is approximately $0$, the minimum of the high
temperature effective potential.  Provided the zero temperature
potential has a sufficiently high barrier separating the metastable
state $\varphi = 0$ from the global minimum (compared to the energy
density in thermal fluctuations at $T = T_c$), then $\varphi
(\undertext{x})$ will remain trapped at $\varphi = 0$ also for $T <
T_c$.  In the notation of Ref. 52, the field $\varphi$ is trapped in
the false vacuum.  After some time (determined again by the potential
barrier), the false vacuum will decay by quantum tunnelling.
\par
Tunnelling in quantum field theory was discussed in Refs. 52-55 (for
reviews see e.g., Refs. 56 and 8).  The transition proceeds by bubble
nucleation.  There is a probability per unit time and volume that at a
point $\undertext{x}$ in space a bubble of ``true vacuum" $\varphi
(\undertext{x}) = a$ will nucleate.  The nucleation radius is
microscopical.  As long as the potential barrier is large, the bubble
radius will increase with the speed of light after nucleation.  Thus,
a bubble of $\varphi = a$ expands in a surrounding ``sea" of false
vacuum $\varphi = 0$.
\par
To conclude, let us stress the most important differences between the
two types of phase transitions discussed above. In a second order
transition, the dynamics is determined mainly by
classical physics.  The transition occurs homogeneously in space
(apart from the phase boundaries which -- as discussed below -- become
topological defects), and $< \varphi (x) >$ evolves continuously in
time.  In first order transitions, quantum mechanics is essential.
The process is extremely inhomogeneous, and $< \varphi (x) >$ is
discontinuous as a function of time.  As we shall see in the following
sections, the above two types of transitions are the basis of various
classes of inflationary Universe models.

\section{Old Inflation}
\par
The old inflationary Universe model$^{13,57)}$ is based on a scalar field
theory which undergoes a first order phase transition.  As a toy
model, consider a scalar field theory with the potential $V (\varphi)$
of Example B (see Fig. 11).  Note that this potential is fairly general
apart from the requirement that $V (a) = 0$, where $\varphi = a$ is
the global minimum of $V (\varphi)$.  This condition is required to
avoid a large cosmological constant today (no inflationary Universe
model manages to avoid the cosmological constant problem).
\par
For fairly general initial conditions, $\varphi (x)$ is trapped in the
metastable state $\varphi = 0$ as the Universe cools below the
critical temperature $T_c$.  As the Universe expands further, all
contributions to the energy-momentum tensor $T_{\mu \nu}$ except for
the contribution
$$
T_{\mu \nu} \sim V(\varphi) g_{\mu \nu} \eqno\eq
$$
redshift.  Hence, the equation of state approaches $p = - \rho$, and
inflation sets in.  Inflation lasts until the false vacuum decays.
During inflation, the Hubble constant is given by
$$
H^2 = {8 \pi G\over 3} \, V (0) \, . \eqno\eq
$$
\par
\midinsert \vskip 6cm
\hsize=6in \raggedright
\noindent {\bf Figure 12.} A sketch of the spatially inhomogeneous
distribution of $\varphi$ in Old Inflation.
\endinsert
After a period $\Gamma^{-1}$, where $\Gamma$ is the tunnelling decay
rate, bubbles of $\varphi = a$ begin to nucleate in a sea of false
vacuum $\varphi = 0$.  For a sketch of the resulting inhomogeneous
distribution of $\varphi (x)$ see Fig. 12.  Note that inflation stops
after bubble nucleation.
\par
The time evolution in old inflation is summarized in Fig. 13.  We denote the
beginning of inflation by $t_i$ (here $t_i \simeq t_c$), the end by $t_R$
(here $t_R \simeq t_c + \Gamma^{-1}$).
\midinsert \vskip 4cm
\hsize=6in \raggedright
\noindent {\bf Figure 13.}  Phases in the old and new inflationary Universe.
\endinsert
\par
It was immediately realized that old inflation has a serious ``graceful exit"
problem$^{13,58)}$.  The bubbles nucleate after inflation with radius $r \ll
2t_R$ and would today be much smaller than our apparent horizon.  Thus, unless
bubbles percolate, the model predicts extremely large inhomogeneities inside
the Hubble radius, in contradiction with the observed isotropy of the
microwave background radiation.
\par
For bubbles to percolate, a sufficiently large number must be produced so that
they collide and homogenize over a scale larger than the present Hubble
radius.  However, with exponential expansion, the volume between bubbles
expands
exponentially whereas the volume inside bubbles expands only with a low power.
This prevents percolation$^{13,58)}$.  If the Universe were expanding
according to a power law with power $> 1$ (generalized inflation), then
percolation might be possible (see Section 3.7).
\section{New Inflation}
\par
Because of the graceful exit problem, old inflation never was considered to be
a viable cosmological model.  However, soon after the seminal paper by
Guth$^{13)}$, Linde and independently Albrecht and Steinhardt$^{50)}$ put
forwards a modified scenario, the New Inflationary Universe (see also Ref.
59).
\par
\midinsert \vskip 6cm
\hsize=6in \raggedright
\noindent {\bf Figure 14.} A sketch of the spatial distribution of
$\varphi$ in New Inflation after the transition. The symbols $+$ and
$-$ indicate regions where $\varphi = + \sigma$ and $\varphi = -
\sigma$ respectively.
\endinsert
Starting point is a scalar field theory with a double well potential which
undergoes a second order phase transition (Fig. 10).  $V(\varphi)$ is
symmetric and $\varphi = 0$ is a local maximum of the zero temperature
potential.  Once again, it was argued that finite temperature effects confine
$\varphi(\undertext{x})$ to values near $\varphi = 0$ at temperatures $T \ge
T_c$.  For $T < T_c$, thermal fluctuations trigger the instability of $\varphi
(\undertext{x}) = 0$ and $\varphi (\undertext{x})$ evolves towards $\varphi =
\pm \sigma$ by the classical equation of motion
$$
\ddot \varphi + 3 H \dot \varphi - a^{-2} \bigtriangledown^2 \varphi = -
V^\prime (\varphi)\, . \eqno\eq
$$
\par
The transition proceeds by spinodal decomposition (see Fig. 14) and hence
$\varphi(\undertext{x})$ will be homogeneous within a correlation length.  The
analysis will be confined to such a small region.  Hence, in Eq.
(3.21)
we can
neglect the spatial gradient terms.  Then, from (3.6) we can read off the
induced equation of state.  The condition for inflation is
$$
\dot \varphi^2 \ll V (\varphi)\, ,\eqno\eq
$$
\ie~ slow rolling.
\par
Often, the  ``slow rolling" approximation is made to find solutions of
(3.21).
This consists of dropping the $\ddot \varphi$ term.  In this case,
(3.21)
becomes
$$
3 H \dot \varphi \, = - \, V^\prime (\varphi)\, . \eqno\eq
$$
As an example, consider a potential which for $|\varphi| < \sigma$ has the
following expansion near $\varphi = 0$
$$
V (\varphi) = V_0 - {1\over 2} m^2 \varphi^2\, . \eqno\eq
$$
With the above $V (\phi)$, (3.23) has the solution
$$
\varphi (t) \, = \, \varphi (0) \exp \left({m^2\over{3H}} t \right)\eqno\eq
$$
(taking $H = \rm const$ which is a good approximation).  Thus, provided $m \ll
\sqrt{3} H \, , \, \ddot \varphi$ is indeed smaller than the other terms in
(3.21) and the slow rolling approximation seems to be satisfied.
\par
However, the above conclusion is premature$^{60)}$.  Equation (3.21) has a
second solution.  For $m > H$ the solution is
$$
\varphi (t) \simeq \varphi (0) e^{mt}\eqno\eq
$$
and dominates over the previous one.  This example shows that the slow rolling
approximation must be used with caution.  Here, however, the conclusion
remains that provided $m \ll H$, then the model produces enough inflation to
solve the cosmological problems.
\par
There is no graceful exit problem in the new inflationary Universe.  Since the
spinodal decomposition domains are established before the onset of inflation,
any boundary walls will be inflated outside the present Hubble radius.
\par
The condition $m^2 \ll 3 H^2$ which must be imposed in order to obtain
inflation, is a fine tuning of the particle physics model -- the first sign of
problems with this scenario.  Consider \eg~ the model (3.14).  By expanding $V
(\varphi)$ about $\varphi = 0$ we can determine both $H$ and $m$ in terms of
$\lambda$ and $\sigma$.  In order that $m^2 < 3 H^2$ be satisfied we need
$$
\sigma > \, \left( {1 \over{6 \pi}}\right)^{1/2} m_{pl} \, , \eqno\eq
$$
which is certainly an unnatural constraint for models motivated by particle
physics.
\par
Let us, for the moment, return to the general features of the new inflationary
Universe scenario.  At the time $t_c$ of the phase transition, $\varphi (t)$
will start to move from near $\varphi = 0$ towards either $\pm \sigma$ as
described by the classical equation of motion, \ie~ (3.25).  At or soon after
$t_c$, the energy-momentum tensor of the Universe will start to be dominated
by $V(\varphi)$, and inflation will commence.  $t_i$ shall denote the time of
the onset of inflation.  Eventually, $\phi (t)$ will reach large values for
which (3.24) is no longer a good approximation to $V (\varphi)$ and for which
nonlinear effects become important.  The time at which this occurs is $t_B$.
For $t > t_B \, , \, \varphi (t)$ rapidly accelerates, reaches $\pm \sigma$,
overshoots and starts oscillating about the global minimum of $V (\varphi)$.
The amplitude of this oscillation is damped by the expansion of the Universe
and (predominantly) by coupling of $\varphi$ to other fields.  At time $t_R$,
the energy in $\varphi$ drops below the energy of the thermal bath of
particles produced during the period of oscillation.
\par
The evolution of $\varphi (t)$ is sketched in Fig. 15a.  The time period
between $t_B$ and $t_R$ is called the reheating period and is usually short
compared to the Hubble expansion time.  The time evolution of the temperature
$T$ of the thermal radiation bath is sketched in Fig. 15b.
\topinsert \vskip 6cm
\hsize=6in \raggedright
\noindent {\bf Figure 15.}  Evolution of $\varphi (t)$ and $T (t)$ in the new
inflationary Universe.
\endinsert
\par
Reheating in inflationary Universe models has been considered in Refs.
61-63.
One way to view the process is as follows$^{63)}$.  Consider a second scalar
field $\psi$ coupled to $\varphi$ via the interaction Lagrangian
$$
{\cal L}_I = {1\over 2} g \varphi^2 \psi^2\, . \eqno\eq
$$
Then, an oscillating $\varphi (t)$ will act as a time dependent mass with
periodic variations in the equation of motion for $\psi$
$$
\ddot \psi_k + 3 H \dot \psi_k + \left(m_\psi^2 + k^2 a^{-2} (t) + g \varphi^2
(t) \right) \psi_k = 0\eqno\eq
$$
where we have neglected nonlinear terms and expanded $\psi$ into Fourier modes
$\psi_k$.  If the expansion of the Universe can be neglected and for periodic
$\varphi^2 (t)$, the above is the well known$^{64)}$ Mathieu equation whose
solutions have instabilities for certain values of $k$.  These instabilities
correspond to the production of $\psi$ particles with well determined
momenta.  These particles eventually equilibrate and
regenerate a thermal bath.
\par
For $t > t_R$, the Universe is again radiation dominated.  Hence, the stages
of the new inflationary Universe are the same as for old inflation
(Fig. 13).
However, there is a useful order of magnitude relation between scale symmetry
breaking $\sigma$ and $H$.  From
$$
H^2 = \, {8 \pi G\over 3} \, V(0)\eqno\eq
$$
and from the form of the potential (see (3.14)) it follows that
$$
\left({H\over \sigma}\right) \sim \, \lambda^{1/2} \,
\left({\sigma\over{m_{pl}}}\right)\, .\eqno\eq
$$
In particular, for $\sigma \sim 10^{15}$GeV (typical scale of grand
unification) and $\lambda \sim 1$ we obtain $H \sim 10^{11}$GeV.
\par
The new inflationary Universe model -- although it was for a long time
presented as a viable model -- suffers from severe fine tuning and initial
condition problems.  In (3.27) we encountered the first of these problems: in
order to obtain enough inflation, the potential must be fairly flat near
$\varphi = 0$.  A more severe problem will be derived in Chapter 4:
Inflationary Universe models generate energy density perturbations.  The
steeper the potential, the larger the density perturbations.  For a potential
which near $\varphi = 0$ or $\varphi = H$ has the following expansion
$$
V (\varphi) = V (0) - \lambda \, \varphi^4\, ,\eqno\eq
$$
the density perturbations conflict with observations unless
$$
\lambda < 10^{-12}\, .\eqno\eq
$$
\par
This in itself is an unexplained small number problem.  However, even if we
were
willing to accept this we would run into initial condition
problems$^{14,49)}$.  For the new inflationary Universe to proceed in the way
outlined above, it is essential that the field $\varphi$ be in thermal
equilibrium with other fields.  This implies that the constant $g$ coupling
$\varphi$ to other fields should not be too small. However,  a coupling term of
the form (3.28) induces
one loop quantum corrections to the self coupling constant $\lambda$
of the order $g^2$.  Hence, the constraint $\lambda < 10^{-12}$ implies a
constraint $g < 10^{-6}$.  Thus, $\varphi$ will not be in thermal equilibrium
at $t_c$, and hence there will be no thermal forces which localize $\varphi$
close to $\varphi = 0$.
\par
Note that the above problem is not an artifact of using quartic potentials
such as (3.32).  Similar constraints would arise in other (\eg~ quadratic)
models.  However, (3.32) was long considered to be the prototypical shape of
$V (\varphi)$ for small values of $\varphi$ since it is the shape which arises
in Coleman-Weinberg$^{65)}$ models.
\par
In the absence of thermal forces which constrain $\varphi$ to start close to
$\varphi = 0$, the only constraints on $\varphi -$ at least using classical
physics alone -- come from energetic considerations.  Obviously, it is
unnatural to assume that at the initial time $t_i$ the energy density in
$\varphi$ exceeds the energy density of one degree of freedom of the thermal
bath at time $t_i$ (temperature $T_i$).  This implies
$$
\eqalign{V (\varphi (\undertext{x}, t_i)) & < {\pi^2\over{30}} T_i^4\cr
| \bigtriangledown \varphi (\undertext{x}, t_i) |^2 & < {\pi^2\over{30}} T_i^4
a^2 (t_i)\cr
| \dot \varphi^2 (\undertext{x}, t_i) |^2 &< {\pi^2\over{30}} T_i^4}\eqno\eq
$$
In particular, for the double well potential of (3.14), (3.34) implies that
$\varphi (\undertext{x}, t_i)$ can be of the order
$$
\varphi (\undertext{x}, t_i) \sim \lambda^{-1/4} T_i\eqno\eq
$$
which for $T_i > \sigma$ is much larger than $\sigma$.  In a weakly coupled
model, the only natural time to impose initial conditions on $\varphi
(\undertext{x})$ is the Planck time, \ie~ $T_i \sim m_{pl}$.  Hence, the
initial conditions allow and in fact suggest
$$
\varphi (\undertext{x}, t_i) \sim \lambda^{-1/4} m_{pl} \, \gg \, m_{pl} \quad
\rm{for} \> T_i \sim T_{pl},\eqno\eq
$$
These observations lead to the chaotic inflation scenario, the only of the
original inflationary Universe models which can still be considered as a
viable scenario today.
\section{Chaotic Inflation}
\par
Chaotic inflation$^{14)}$ is based on the observation that for weakly coupled
scalar fields, initial conditions which follow from classical considerations
alone lead to very large values of $\varphi (\undertext{x})$ (see
(3.35)).
\par
Consider a region in space where at the initial time $\varphi (\undertext{x})$
is very large, homogeneous (we will make these assumptions quantitative
below) and static.  In this case, the energy-momentum tensor will be
immediately dominated by the large potential energy term and induce an
equation of state $p \simeq - \rho$ which leads to inflation.  Due to the
large Hubble damping term in the scalar field equation of motion, $\varphi
(\undertext{x})$ will only roll very slowly towards $\varphi = 0$.  The
kinetic energy contribution to $T_{\mu \nu}$ will remain small, the spatial
gradient contribution will be exponentially suppressed due to the expansion of
the Universe, and thus inflation persists.  This is a brief survey of the
chaotic inflation scenario.  Note that in contrast to old and new inflation,
no initial thermal bath is required.  Note also that the precise form of
$V(\varphi)$ is irrelevant to the mechanism.  In particular, $V(\varphi)$ need
not be a double well potential.  This is a significant advantage, since for
scalar fields other than Higgs fields used for spontaneous symmetry breaking,
there is no particle physics motivation for assuming a double well potential,
and since the inflaton (the field which gives rise to inflation) cannot be a
conventional Higgs field due to the severe fine tuning constraints.
\par
Let us consider the chaotic inflation scenario in more detail.  For
simplicity, take the potential
$$
V (\varphi) = \, {1\over 2} \, m^2 \varphi^2\eqno\eq
$$
and consider a region in space in which $\varphi (\undertext{x}, t_i)$ is
sufficiently homogeneous.  To be specific, we require
$$
{1\over 2} \, a^{-2} (t_i) | \bigtriangledown \varphi \, (\undertext{x},
t_i)|^2 \, \ll \, V\left(\varphi (\undertext{x}, t_i)\right)\eqno\eq
$$
over a region of size $d_i$
$$
d_i \ge 3 H^{-1} (t_i)\, .\eqno\eq
$$
We also require that the kinetic energy be negligible at the initial time
$t_i$,
$$
\dot \varphi (\undertext{x}, t_i)^2 \, \ll \,
V (\varphi \left(\undertext{x}, t_i) \right)\, ,\eqno\eq
$$
although this assumption can be relaxed without changing the
results$^{66)}$.
 From (3.38) and (3.40) it follows that at $t_i$ the equation of state is
inflationary, \ie~ $p (t_i) \simeq - \rho (t_i)$.  Condition (3.39) ensures
that no large inhomogeneities can propagate to the center of the region under
consideration from outside.  With these approximations, the equation of motion
for $\varphi$ becomes
$$
\ddot \varphi + 3 H \dot \varphi = - m^2 \varphi\eqno\eq
$$
with
$$
H = \left({4 \pi\over 3}\right)^{1/2} \, {m\over{m_{pl}}} \, \varphi\, .
\eqno\eq
$$
\par
Since we expect $\varphi (\undertext{x}, t)$ to be changing slowly, we make
the slow rolling approximation
$$
3 H \dot \varphi = - m^2 \varphi\eqno\eq
$$
which gives
$$
\dot \varphi = - \left({1\over{12 \pi}}\right)^{1/2} m \, m_{pl}\eqno\eq
$$
and shows that the approximation is self consistent.  In order to get
inflation, we require
$$
{1\over 2} \, \dot \varphi^2 < {1\over 2} m^2 \, \varphi^2\eqno\eq
$$
which (by (3.44)) is satisfied if
$$
\varphi > \left({1\over{12}}\right)^{1/2} m_{pl}\, . \eqno\eq
$$
In order to obtain a period $\tau > 50 H^{-1} (t_i)$ of inflation, a slightly
stronger condition is needed$^{14)}$:
$$
\varphi > 3 m_{pl}\, . \eqno\eq
$$
\par
With chaotic inflation, the initial hope that grand unified theories could
provide the answer to the homogeneity and flatness problems has been
abandoned.  The inflaton is introduced as a new scalar field (with no
particular particle physics role which is very weakly coupled to itself and to
other fields (see Ref. 67 for an attempt to couple the inflaton to non-grand
unified particle physics).  In supergravity and in superstring inspired models
there are scalar fields which are candidates to be the inflaton.  I refer the
reader to Refs. 68-70 for a discussion of this issue.  However, even in such
models the time when the inflaton $\varphi$ decouples from the rest of physics
is the Planck time $t_i = t_{pl}$.  Thus, the chaotic inflation scenario is
often called primordial inflation.
\par
Chaotic inflation predicts that the Universe is locally flat and homogeneous,
but globally extremely curved and inhomogeneous$^{71)}$.  This is in marked
contrast to new inflation where the Universe is predicted to be globally
homogeneous.
\par
Chaotic inflation is a much more radical departure from standard cosmology
than old and new inflation.  In the latter, the inflationary phase can be
viewed as a short phase of exponential expansion bounded at both ends by
phases of radiation domination.  In chaotic inflation, a piece of the Universe
emerges with an inflationary equation of state immediately after the quantum
gravity epoch.
\section{Alternative Models for Inflation}
\par
Old inflation, new inflation and chaotic inflation are all based on the use of
new fundamental scalar fields which cannot be the Higgs fields of any known or
grand unified gauge theory.  Instead of introducing new physics to give
inflation via scalar fields, it is possible to look for realizations of
inflation with some alternative new physics which does not use scalar
fields.  In particular, it is possible to consider modifications of
Einstein gravity.
\subsection{{\bf a)} ~ $R^2$ gravity}
\par
The original attempt at obtaining inflation from other sources than from
scalar fields in fact predates inflation.  It is the Starobinsky
model$^{72)}$
in which the Lagrangian for gravity is modified by including a term
proportional to $R^2$.  In the initial model$^{72)}$, this term was introduced
as a contribution to the effective action from quantum corrections.  Later,
a model with bare Lagrangian including $\epsilon R^2$ was
studied$^{73)}$.
\par
Consider the theory with Lagrangian for gravity
$$
{\cal L} = R + \epsilon R^2\eqno\eq
$$
where $R$ is the Ricci scalar of a metric $g_{\mu \nu}$.  As shown by
Whitt$^{74)}$, the equations of motion which result from this Lagrangian are
the same as those for a system with metric
$$
\tilde g_{\mu \nu} = ( 1 + 2 \epsilon R) g_{\mu \nu}\eqno\eq
$$
and scalar field $R$. The time evolution of the metric $\tilde g_{\mu \nu}$
is described by the Lagrangian
$$
\tilde {\cal L} = \tilde R \, ,\eqno\eq
$$
where $\tilde R$ is the Ricci scalar curvature for the metric $\tilde
g_{\mu\nu}$,
and that of the scalar field
$$
\varphi \equiv \ln (1 + 2 \epsilon R) \eqno\eq
$$
by the usual scalar field action with potential
$$
V (\varphi) = - {1\over{4 \varepsilon}} (1 - e^{-\varphi})^2 \eqno\eq
$$
Speaking very qualitatively, we can view the above as the action of a scalar
field which for $\epsilon \gg 1$ (in Planck units) is very weakly coupled and
which has a quadratic potential.  Hence, provided $R(t_i)$ is sufficiently
large, chaotic-type inflation results.
\par
The fact that higher derivative corrections to Einstein gravity lead
to inflation is not specific to the quadratic Lagrangian (3.48).  The
case of a general Lagrangian ${\cal L} = f (R)$ has been
discussed in Ref. 15.  It can also be analyzed simply using a
conformal transformation.  The ``nonsingular Universe" model of Ref.
17 also leads to inflation at high curvatures driven by higher
derivative terms in the gravitational action.

\subsection{{\bf c)}~ Extended inflation}
\par
Extended inflation$^{75,76)}$ is a revival of old inflation in the context of a
Brans-Dicke theory of gravity.  The action is
$$
S = \int \, d^4 x \, \sqrt{-g} \, \left( \varphi R - \omega \, {(\partial_\mu
\varphi)^2\over \varphi} + \, 16 \pi {\cal L}_m \right)\, . \eqno\eq
$$
Here, $\varphi$ is the Brans-Dicke scalar field (which induces a time and
space dependent
gravitational coupling constant), $\omega$ is the Brans-Dicke parameter and
${\cal L}_m$ is the Lagrangian for matter fields.  ${\cal L}_m$ is chosen to
give rise to a first order phase transition which proceeds by nucleation of
true vacuum bubbles in a surrounding sea of false vacuum.
\par
The crucial point is that in Brans-Dicke theory, an equation of state $p = -
\rho$ for matter leads to power law rather than exponential expansion of the
Universe.  For a FRW ansatz for $g_{\mu \nu}$ in terms of a homogeneous scale
factor $a(t)$ and homogeneous $\varphi (t)$, the equations of motion
are (for $k=0$)
$$
H^2 = \, {8 \pi \rho\over{3 \varphi}} \, + \, {\omega\over 6} \, \left({\dot
\varphi\over \varphi}\right)^2 - H \, {\dot \varphi\over \varphi}\eqno\eq
$$
and
$$
\ddot \varphi + 3 H \dot \varphi = \, {8 \pi (\rho - 3p)\over{3 + 2 \omega}}
\, . \eqno\eq
$$
for $p = - \rho$, the solution is
$$
\varphi (t) = \, m_{pl}^2 (1 + Ht/\alpha)^2\eqno\eq
$$
and
$$
a (t) = (1 + Ht/\alpha)^{\omega + 1/2}\eqno\eq
$$
with $\alpha \sim \omega^2$ and
$$
H^2 = \, {8 \pi\over 3} \, m_{pl}^{-2} \rho \, . \eqno\eq
$$
\par
Since the false vacuum only expands as a power of $t$ in Brans-Dicke theory,
true vacuum bubbles are able to percolate$^{76)}$.  The probability of a point
remaining in the false vacuum is$^{58)}$
$$
p (t) = \exp \left[ - \int\limits^t_{t_c} dt^\prime \, \lambda (t^\prime) a
(t^\prime)^3 \, {4 \pi\over 3} \left( \int\limits^t_{t^\prime} \,
{dt^{\prime\prime}\over{a(t^{\prime\prime})}} \right)^3 \right]\eqno\eq
$$
where $\lambda (t)$ is the nucleation rate per unit time and $t_c$ is the
critical time of the phase transition (when nucleation begins).  With
$\epsilon = \, {\lambda\over{H^4}}$
taken to be time independent, we obtain
$$
p (t) \sim \exp \left[ - {4 \pi\over 3} \, \epsilon H\right]\, . \eqno\eq
$$
\par
In the old inflationary Universe, $p (t) a^3 (t)$ diverges since $\epsilon \ll
1$ and $a(t)$ increases exponentially.  This implies that true vacuum bubbles
do not percolate and we can never produce a region the size of our Universe
with true vacuum.  However, in extended inflation, $a (t)$ grows as a power of
$t$.  Hence
$$
\lim_{t \rightarrow \infty} a^3 (t) p (t) = 0\eqno\eq
$$
and true vacuum bubbles percolate.
\par
However, percolation of true vacuum bubbles is not sufficient to give a viable
cosmological model.  The bubbles must also thermalize in order not to produce
too large density perturbations.  According to a recent
analysis$^{77)}$,
thermalization occurs only if $\omega < 10$, a bound which conflicts with the
lower bound $\omega > 500$ stemming from time delay measurements.  Hence, in
order to save extended inflation, new variants of the models are required such
as$^{78)}$ a model with a nontrivial potential $V(\varphi)$ for the Brans-
Dicke scalar.
\section{Concluding Comments}
\par
As has hopefully become clear in this and the previous chapter, inflation
is a nice idea which solves many problems of standard big bang
cosmology.  However, no convincing realization of inflation which does
not involve unexplained small numbers has emerged (for a general
discussion of this point see Ref. 79).
\par
It is important to distinguish between models of inflation which are
self consistent and those which are not.  We have shown that new
inflation is not self consistent, whereas chaotic and improved
versions of extended inflation are.  One of the key issues involves
initial conditions.  In new inflation, the initial conditions required
can only be obtained if the inflaton field is in thermal equilibrium
above the critical temperature, which however is not possible because
of the density fluctuation constraints on coupling constants.
\par
In chaotic inflation, it can be shown that$^{80-82)}$ -- provided the
spatial sections are flat -- a large phase space of initial conditions
(much larger than is apparent from (3.38) and (3.40)) gives chaotic
inflation, whereas the probability to relax dynamically$^{83)}$ to
field configurations which give new inflation (this possibility is
only available in double well potentials) is negligibly small.

\chapter{Fluctuations in Inflationary Universe Models}
\section{Preliminaries}
\par
In this chapter, the origin of the primordial density perturbations required
to seed galaxies will be discussed within the context of inflationary Universe
models.  From Chapter 2, we recall the basic reason why in inflationary
cosmology a causal generation mechanism is possible: comoving scales of
cosmological interest today originate inside the Hubble radius early in the de
Sitter period (see Fig. 8).  Hence, it is in principle possible that density
perturbations on these scales can be generated by a causal mechanism at very
early times.
\par
First, let us demonstrate why the Hubble radius $H^{-1} (t)$ is the length
scale of relevance in these considerations.  Consider a scalar field theory
with action
$$
S (\varphi) = \int d^4 x \sqrt{-g} \, \left[ {1\over 2} \partial_\mu \varphi
\partial^\mu \varphi - V (\varphi) \right] \, . \eqno\eq
$$
The resulting equation of motion is
$$
\ddot \varphi + 3 H \dot \varphi - a^{-2} \nabla^2 \varphi = - V^\prime
(\varphi) \, . \eqno\eq
$$
The second term on the left hand side is the Hubble damping term, the third
represents microphysics (spatial gradients).  To simplify the consideration,
assume that $V^\prime (\varphi) = 0$.  Then, the time evolution is influenced
by microphysics and gravity.  For plane wave perturbations with wave number
$k$, the gravitational force is proportional to $H^2 \varphi$ whereas the
microphysical force is $a^{-2} k^2 \varphi$.  Thus, for $ak^{-1} < H^{-1}$,
i.e., on length scales smaller than the Hubble radius, microphysics dominates,
whereas for $ak^{-1} > H^{-1}$, i.e., on length scales larger than the Hubble
radius, the gravitational drag dominates.
\par
Based on the above analysis we can formulate the main idea of the fluctuation
analysis in inflationary cosmology.  In linear order, all Fourier
modes
decouple.  Hence, we fix a mode with wave number $k$.  There are two very
different time intervals to consider.  Let $t_i (k)$ be the time when the
scale crosses the Hubble radius in the de Sitter phase, and $t_f (k)$ the time
when it reenters the Hubble radius after inflation (see Fig. 16).  The first
period runs for $t < t_i (k)$.  In this time interval, microphysics
dominates.  We shall demonstrate that quantum fluctuations generate
perturbations during this period.  The second time interval is $t_i (k) < t <
t_f (k)$.  Now, microphysics is unimportant and the evolution of perturbations
is determined by gravity.  At $t_i (k)$, decoherence sets in; the quantum
mechanical wave functional can be viewed as a statistical ensemble of
classical configurations after this time.
\par
\midinsert \vskip 7cm
\hsize=6in \raggedbottom
\noindent {\bf Figure 16.} Sketch of the evolution of two fixed
comoving scales in the inflationary Universe.
\endinsert
In this chapter, I will first discuss the classical evolution of fluctuations
outside the Hubble radius and the technical issues of coordinate invariance
which must be addressed.  The second topic is the quantum generation of
perturbations for $t < t_i (k)$.  Section 4.4 contains an analysis of
decoherence based on Ref. 84.
\par
I choose to present the material in three distinct steps to highlight the
various deep physical questions involved.  Once these issues are understood, a
unified treatment is possible (see Ref. 15).
\par
As a last preliminary, let us introduce the necessary notation.  The quantity
of interest is the r.m.s. mass excess $(\delta M/M) \, (k, t)$ at time $t$ in
a sphere of radius $k^{-1}$.  Given a smooth density distribution
$$
\rho ( \undertext{x} , t) = \rho_0 (t) + \delta \rho ( \undertext{x} , t) \, ,
\eqno\eq
$$
the r.m.s. mass excess can be related to the Fourier mode $\delta \rho (k)$ in
a straightforward manner$^{9)}$.  The result is
$$
\left( {\delta M\over M} \right)^2 \, (k, t) \simeq k^3 \big| {\delta
\rho\over \rho_0} \big|^2 \, (k,t) \, . \eqno\eq
$$
The adopted convention for Fourier transformation is
$$
\delta \rho (\undertext{x}) = (2 \pi)^{- 3/2} V^{-1/2} \int d^3 k e^{i
\undertext{k} \undertext{x}} \delta \rho (\undertext{k}) \, . \eqno\eq
$$
The quantity $(\delta \rho/ \rho_0)^2 (k)$ is called the power spectrum.  The
result (4.4) holds provided the power spectrum is proportional to $k^n$ with
$n > -3$.  An intuitive way to understand the result is as follows:
perturbations with wave number larger than $k$ average to zero in a volume
$k^{-3}$, perturbations with wave number smaller then $k$ are phase space
suppressed such that $(\delta M / M) \, (k)$ receives its major contribution
from Fourier modes of wave number $k$.  Their phase space volume is $k^3$.
\section{Gauge Invariance and Classical Evolution}
\par
By the Einstein equations, a nonvanishing density perturbation implies the
existence of metric fluctuations.  The full metric $g_{\mu \nu}$ can be
decomposed into a space-independent background piece $g_{\mu \nu}^{(0)} (t)$
and a perturbation $\delta g_{\mu \nu} (\undertext{x} , t)$:
$$
g_{\mu \nu} (\undertext{x} , t) = g_{\mu \nu}^{(0)} (t) + \delta g_{\mu \nu}
(\undertext{x}, t) \, . \eqno\eq
$$
However, this splitting depends on the choice of coordinates $\undertext{x}$
and $t$.  Physical quantities must be independent of the choice of
coordinates.  Hence, not all components of $\delta g_{\mu \nu} (\undertext{x} ,
t)$ can be physical.
\par
\midinsert \vskip 10cm
\hsize=6in \raggedbottom
\noindent {\bf Figure 17.} Sketch of how two choices of the mapping
between the background space-time manifold ${\cal M}_0$ to physical
space-time ${\cal M}$ induce two different coordinate systems on
${\cal M}$.
\endinsert
A change in coordinates is called a gauge transformation.  Addressing the
gauge dependence of $\delta g_{\mu \nu} (\undertext(x), t)$ is a key issue in
the study of cosmological perturbations.  To illustrate this point, consider
Fig. 17.  A choice of coordinates on the physical manifold ${\cal M}$ can be
viewed as a mapping ${\cal D}$ between a background manifold ${\cal M}_0$
(with fixed coordinates) and ${\cal M}$.  Consider a physical quantity $Q$ on
${\cal M}$ and define a background function $Q^{(0)}$ on ${\cal M}_0$.  Then,
in the coordinate system given by ${\cal D}$, the quantity $Q$ can be
decomposed into background and perturbation $\delta Q$ as
$$
\delta Q (p) = Q (p) - Q^{(0)} ({\cal D}^{-1} (p) ) \eqno\eq
$$
where $p$ is a point on ${\cal M}$.  In a different coordinatization given by
a second map $\tilde {\cal D}$
$$
\delta \tilde Q (p) = Q (p) - Q^{(0)} (\tilde D^{-1} (p) ) \, . \eqno\eq
$$
The difference $\delta Q (p) - \delta \tilde Q (p)$ is obviously a gauge
artifact.  Similarly, a change in ${\cal D}$ leads to a change in $\delta
g_{\mu \nu}$.
\par
There are two ways to address the gauge freedom.  It is possible to choose a
particular gauge and to only work in this gauge.  The problem with this
approach is that the quantities appearing in the calculation have no gauge
invariant (and thus physical) meaning.  In addition, some popular gauge
choices do not totally fix the coordinate freedom.  A better approach is to
develop the theory of cosmological perturbations in terms of gauge invariant
variables.
\par
The gauge invariant theory of cosmological perturbations is in principle
straightforward, although technically rather tedious.  I will in the following
summarize the main steps and refer the reader to Ref. 15 for the
details and
further references (see also Ref. 85 for a pedagogical introduction).
\par
We consider perturbations about a spatially flat Friedmann-Robertson-Walker
metric
$$
ds^2 = a^2 (\eta) (d \eta^2 - d \undertext{x}^2) \eqno\eq
$$
where $\eta$ is conformal time (related to cosmic time $t$ by $a (\eta) d \eta
= dt$).  A scalar metric perturbation (see Ref. 86 for a precise definition)
can be written in terms of four free functions of space and time:
$$
\delta g_{\mu \nu} = a^2 (\eta) \pmatrix{2 \phi & - B_{,i} \cr
-B_{,i} & 2 (\psi \delta_{ij} + E_{,ij} \cr} \, . \eqno\eq
$$
Scalar metric perturbations are the only perturbations which couple to energy
density and pressure.
\par
The next step is to consider infinitesimal coordinate transformations
$$
x^{\mu^\prime} = x^\mu + \xi^\mu \eqno\eq
$$
which preserve the scalar nature of $\delta g_{\mu \nu}$ and to calculate the
induced transformation of $\phi, \psi, B$ and $E$.  Then, we find invariant
combinations to linear order.  (Note that there are in general no combinations
which are invariant to all orders$^{87)}$.)  After some algebra, it follows
that
$$
\eqalign{\Phi & = \phi + a^{-1} [ (B - E^\prime ) a ]^\prime \cr
\Psi & = \psi - {a^\prime\over a} (B - E^\prime) } \eqno\eq
$$
are two invariant combinations.  In the above, a prime denotes differentiation
with respect to $\eta$.
\par
There are various methods to derive the equations of motion for gauge
invariant variables$^{88-93)}$.  Perhaps the simplest way is to consider the
linearized Einstein equations
$$
\delta G_{\mu \nu} = 8 \pi G \delta T_{\mu \nu} \eqno\eq
$$
and to write them out in longitudinal gauge defined by
$$
B = E = 0 \eqno\eq
$$
and in which $\Phi = \phi$ and $\Psi = \psi$, to directly obtain gauge
invariant equations$^{15)}$.
\par
For several types of matter, in particular for scalar field matter, the
perturbation of $T_{\mu \nu}$ has the special property
$$
\delta T^i_j \sim \delta^i_j \eqno\eq
$$
which implies $\Phi = \Psi$.  Hence, the scalar-type cosmological
perturbations can in this case be described by a single gauge invariant
variable.  The equation of motion takes the form
$$
\dot \xi = O \left( {k\over aH} \right)^2 H \xi \eqno\eq
$$
where
$$
\xi = {2\over 3} \, {H^{-1} \dot \Phi + \Phi\over{1 +w}} + \Phi \, . \eqno\eq
$$
The variable $w = p/\rho$ (with $p$ and $\rho$ background pressure and energy
density respectively) is a measure of the background equation of state.  In
particular, on scales larger than the Hubble radius, the right hand side of
(4.16) is negligible, and hence $\xi$ is constant.
\par
The equations (4.16) and (4.17) can be immediately be applied to the growth of
classical cosmological perturbations outside the Hubble radius, i.e., between
$t_i (k)$ and $t_f (k)$ in inflationary Universe models.  During inflation
$$
1 + w (t_i) = \dot \varphi^2 (t_i)/ \rho \ll 1 \, , \eqno\eq
$$
whereas after inflation (in the radiation dominated epoch)
$$
1 + w (t_f) = {4\over 3} \, . \eqno\eq
$$
Hence, $\Phi$ increases by a large factor between $t_i (k)$ and $t_f (k)$:
$$
\Phi (t_f) \simeq {4\over 9} \, {1\over{1 + w(t_i) }} \, \Phi (t_i) \, .
\eqno\eq
$$
\par
To make the connection with observable quantities, notice that at any Hubble
radius crossing $t_H$, the gravitational potential $\Phi$ is proportional to
the relative energy density perturbation $\delta \rho / \rho_0$ (in a comoving
coordinate system).  Hence, the energy density perturbation increases by the
same large factor between $t_i (k)$ and $t_f (k)$.
\par
Without a careful analysis of the quantum generation process of fluctuations,
it is not possible to precisely calculate $\delta M / M (t_f)$.  However, a
rough estimate can be obtained as follows: the initial density perturbation
$(\delta M/M) (t_i)$ will be determined by the strength of the de Sitter
quantum fluctuations$^{44,47,94)}$, which in turn is given by the Hawking
temperature$^{95)}$ $T_H \sim H$.  Also, $\dot \varphi (t_i) \sim H^2$ since
the
initial motion of $\varphi$ is induced by the Hawking fluctuations.  Hence
$$
{\delta M\over M} \, (t_f (k) ) \sim 1 \, . \eqno\eq
$$
In order to produce density perturbations of the right order of magnitude to
lead to galaxies today, and in order not to generate too large microwave
background anisotropies, a value of the order $10^{-4}$ is required.  The
conflict between (4.21) and the required value has become known as the
fluctuation problem.  The only known solutions to date involve small numbers
(unappealing!) in the particle physics sector.
\section{Quantum Generation of Perturbations}
\par
A subtle issue is the question of the origin of classical density fluctuations
in the de Sitter phase of an inflationary cosmology.  Based on a homogeneous
quantum state $| \psi >$ for matter we want to construct a classical
inhomogeneous energy-momentum tensor $T_{\mu \nu}^{cl} (\undertext{x}, t)$
which couples to gravity.
\par
Quantum fluctuations are the key idea in understanding the origin of classical
density perturbations in inflationary cosmology$^{44,47,94)}$.  The homogeneous
functional for the state $| \psi >$ can be viewed as a statistical
superposition of classical configurations, each of which will in general be
inhomogeneous.  The quantum to classical transition will pick out a ``typical"
configuration and use it to define $T^{cl}_{\mu \nu} (\undertext{x} , t)$.
This procedure makes sense as long as there is decoherence, i.e., minimal
interference between the different trajectories represented in $| \psi >$ (see
Section 4.4).  We shall show that decoherence starts at $t_i (k)$.
\par
This procedure can be implemented as follows$^{96,97)}$:
Define a classical field
$$
\varphi_{cl} (\undertext{x}, t) = \varphi_0 (t) + \delta \varphi
(\undertext{x}, t)\eqno\eq
$$
(with vanishing spatial average of $\delta \varphi$).  Then, $T_{\mu
\nu}^{cl}$ is given by
$$
T_{\mu \nu}^{cl} (\undertext{x}, t) = T_{\mu \nu} \left(\varphi_{cl}
(\undertext{x}, t) \right)\, , \eqno\eq
$$
where $T_{\mu \nu} (\varphi_{cl})$ is defined as the usual energy-momentum
tensor$^{98)}$ of a classical scalar field.  Note that unless $\delta \varphi$
vanishes, $T_{\mu \nu}^{cl}$ is not homogeneous.
\par
Each Fourier mode $\delta \tilde \varphi (\undertext{k}, t)$ of
$\varphi (\undertext{x}, t)$ is a harmonic oscillator. The quantum field
theory ground state $|\psi >$ is the harmonic oscillator ground state for
each $k$.  We can define $\delta \varphi$ in Fourier space by identifying
$\delta \tilde \varphi (\undertext{k})$ with the spread of the ground
state wave function for the oscillator.
$$
| \delta \tilde \varphi (\undertext{k}) |^2 \equiv < \psi | \, |
\tilde \varphi (\undertext{k}) |^2 |\psi >\, . \eqno\eq
$$
This corresponds to picking out the most likely classical
configuration.  Similarly, $\varphi_0 (t)$ is defined by
$$
\varphi_0 (t)^2 \equiv <\psi| \varphi (\undertext{x})^2 | \psi >\eqno\eq
$$
which is position independent since $| \psi >$ is taken to be homogeneous.
Assuming that $\delta \varphi$ is small compared to $\varphi$, to first order
the spatial gradients do not contribute to the energy density perturbation
$\delta \rho$.  In Fourier space,
$$
\delta \tilde \rho (\undertext{k}) = \dot \varphi_0 \delta \dot{\tilde
\varphi}
(\undertext{k}) + V^\prime (\varphi_0) \delta \tilde \varphi (\undertext{k})\,
. \eqno\eq
$$
The above is to be evaluated at $t_i (k)$.
\par
The computation of the spectrum of density perturbations produced in the de
Sitter phase has been reduced to the evaluation of the expectation values
(4.24) and (4.25).  First, we must specify the state $| \psi >$.  (Recall that
in non-Minkowski space-times there is no uniquely defined vacuum state of a
quantum field theory$^{99)}$).  We pick the FRW frame of the pre-inflationary
period.  In this frame, the number density of particles decreases
exponentially.  Hence we choose $| \psi >$ to be the ground state in this
frame (see Ref. 100 for a discussion of other choices).  $\psi [ \tilde \varphi
(\undertext{k}), t]$, the wave functional of $|\psi >$ can be calculated
explicitly$^{97)}$.  It is basically the superposition of the ground state
wave function for all oscillators
$$
\psi [ \tilde \varphi (\undertext{k}), t] = N \exp \left\{ - {1\over 2} (2
\pi)^{-3} a^3 (t) \int d^3 k \omega (\undertext{k}, t) | \tilde \varphi
(\undertext{k})|^2 \right\} \, . \eqno\eq
$$
$N$ is a normalization constant and $\omega (\undertext{k}, t) \sim H$ at $t =
t_i (k)$.  Hence
$$
\delta \tilde \varphi (\undertext{k}, t) = (2 \pi)^{3/2}  a^{-3/2}
\omega (\undertext{k}, t)^{-1/2} \sim (2 \pi)^{3/2}  k^{-3/2} H \, , \,
t = t_i (k)\, . \eqno\eq
$$
Note that $\delta \tilde \varphi (\undertext{k}, t)$ is well defined without
regularization and renormalization.  In contrast, $\varphi_0 (t)$ is not
apriori well defined.  Inserting $\varphi^2 (\undertext{x})$ expressed as an
integral over the Fourier coefficient operators $\tilde \varphi
(\undertext{k})$
into (4.25), we find both an ultraviolet (UV) and an infrared (IR) divergence.
Hence, regularization and renormalization are needed.  We eliminate the IR
divergence by imposing a physical cutoff:$^{101)}$  modes with $|\undertext{k}
a^{-1} (t) |<H$ are taken not to contribute.  They simply modify the
background.  The UV divergence is regulated and renormalized by subtracting
the Minkowski space value.  These prescriptions lead to$^{101,97)}$
$$
\varphi_0 (t) \simeq (2 \pi)^{-1} H^{3/2}\,  t^{1/2}\, . \eqno\eq
$$
\par
Combining (4.26), (4.28) and (4.29) and using $t \sim H^{-1}$ leads to
$$
\delta \tilde \rho \left(\undertext{k}, t_i (\undertext{k}) \right) = O
(1) k^{-3/2} H^4\eqno \eq
$$
as our final result for the spectrum of density perturbations generated by
quantum fluctuations in the de Sitter phase of an inflationary Universe.  In
terms of physical quantities, (see (4.4))
$$
{\delta M\over M} \, \left(\undertext{k}, t_i (\undertext{k}) \right) = O
(1) H^4 \rho^{-1} \, , \eqno\eq
$$
and from (4.20) it follows that
$$
{\delta M\over M} (k, t_f (k) ) = O (1) \, {H^4\over{\dot \phi^2 (t_i (k)
) }} \, \eqno\eq
$$
in agreement with the result of the heuristic analysis at the end of Section
4.2.  Evaluated for a chaotic inflation potential
$$
V (\varphi) = {1\over 4} \lambda \varphi^4 \eqno\eq
$$
we obtain$^{102)}$ -- similar to the case of new inflation$^{103)}$ -- the
result
$$
{\delta M\over M} (k, t_f (k) ) \sim \lambda^{1/2} 10^2 \, . \eqno\eq
$$
Thus, a value $\lambda \sim 10^{-12}$ is required in order to obtain a
reasonable result for the primordial perturbations.
\section{Decoherence}
\par
In order to tie together the quantum generation mechanism for cosmological
perturbations discussed above with the classical analysis of the evolution
summarized in Section 4.2, it is important to demonstrate that cosmological
perturbations decohere on scales larger than the Hubble radius.
\par
In the following, I shall summarize the analysis of Ref. 84 (for earlier work
on this subject see e.g., Refs. 104-106).  As stated before, the key questions
is
to determine when the quantum mechanical wave functional evolves like a
statistical average of classical configurations.  A sufficient condition is
the decoherence property of the density matrix.
\par
The density matrix $\rho (f, \tilde f)$ of a quantum system is said to
decohere if
$$
\rho (f, \tilde f) \simeq 0 \qquad {\rm for} \qquad f = \tilde f \, , \eqno\eq
$$
where $f$ and $\tilde f$ are classical field configurations.  For a pure
quantum state $| \psi >$ with wave functional $\psi (f)$, the density matrix
is
$$
\rho (f, \tilde f) = \psi^\ast (f) \psi (\tilde f) \, , \eqno\eq
$$
and hence the decoherence property (4.35) will not be satisfied.
\par
The key idea$^{105,106)}$ is that decoherence is induced by the interaction of
the system with an environment which is not observed.  In this case, the
important quantity is the reduced density matrix $\rho_{\rm red}$ obtained by
tracing over the environment.  If $f_1$ and $f_2$ denote system and
environment configurations respectively, then the reduced density matrix of
the system is
$$
\rho_{\rm red} (f_1 , \, \tilde f_1) = \int d f_2 \psi^\ast (f_1 , \, f_2)
\psi (\tilde f_1 \, , f_2) \eqno\eq
$$
It is possible that the reduced density matrix decoheres even if $\psi (f_1 ,
\, f_2)$ is a pure state.
\par
To illustrate this effect, consider an exactly solvable toy model consisting
of two derivatively coupled free scalar fields.  Let $\varphi_1$ stand for the
system (the inflaton) and $\varphi_2$ for the environment (other matter
fields, gravitational waves etc.).  The action in an expanding Universe is
$$
S = \int d^4 x \sqrt{-g} \, {1\over 2} \, \left( \partial_\mu \varphi_1 \,
\partial^\mu \varphi_1 + \partial_\mu \varphi_2 \partial^\mu \varphi_2 + 2 c
\partial_\mu \varphi_1 \partial^\mu \varphi_2 \right) \, . \eqno\eq
$$
Note that this system admits a diagonal basis of fields in which there is no
interaction.  However, there is not reason to assume that system and
environment form this basis, and we shall not make this assumption.
\par
However, the existence of the diagonal basis is useful when constructing the
wave functional for the state of the full quantum system.  First, however, we
recall that for the above action there is no coupling between the different
Fourier modes.  Hence we can fix a wave vector $\undertext{k}$ and reduce the
problem to a two degree of freedom quantum mechanics problem.  Let $f_1$ and
$f_2$ be the field coordinates corresponding to the Fourier modes $\tilde
\varphi_1 (k) $ and $\tilde \varphi_2 (k)$.  The vacuum state wave functional
is a product of Gaussians for each harmonic oscillator.  Hence
$$
\psi (f_1 , f_2) = {1\over{(2 \pi)^{1/2} }} \, {1\over g} \exp \, \left\{ i
{a^2\over 2} \, \left( {g^\prime\over g} \right) \, (f^2_1 + f^2_2 + 2c f_1
f_2)
\right\} \, ,  \eqno\eq
$$
where a prime denotes the derivative with respect to conformal time $\eta$,
and $g (\eta)$ is the mode function satisfying the equation
$$
(a^2 g^\prime)^\prime + a^2 k^2 g = 0 \, . \eqno\eq
$$
The ambiguity in the solution corresponds to the well-known ambiguity in the
choice of the vacuum state in an expanding Universe.  Choosing the
Bunch-Davies$^{99)}$ vacuum, and after some algebra, we obtain the
result
$$
\rho (x, y ; \eta) = N \exp \left\{ - \left( \left(x\over \sigma \right)^2 +
\left( y\over \ell_c \right)^2 + 2 i \theta x y \right) \right\} \, ,
\eqno\eq
$$
where the normalization constant $N$ and the real phase constant $\theta$ are
irrelevant for the conclusions.  In the above, we have introduced the field
variables
$$
\eqalign{ x & = {1\over 2} (f_1 \, + \tilde f_1 ) \cr
y & = {1\over 2} (f_1 \, - \tilde f_1) \, . } \eqno\eq
$$
The constants $\sigma$ and $\ell_c$ in (4.41) are called dispersion and
coherence length respectively.  Note that these are both lengths in
configuration space.
\par
The coherence length $\ell_c$ depends on $k$ and $\eta$ in the following way
$$
\ell_c = {H\over{k^{3/2} }} \> {[(k\eta)^2 + 1 ]^{1/2}\over{[1+ c^2/(k \eta)^2
] }} \, . \eqno\eq
$$
Note that the Hubble radius crossing criterium is $(k \eta)^2 = 1$.  Thus,
from (4.43) it follows that for $c \neq 0$ (nonzero coupling between system
and environment) the coherence length tends to 0 as the wavelength grows
larger than the Hubble radius $(k \eta \rightarrow 0)$.  Hence, as is obvious
from (4.41), decoherence sets in once the scale becomes larger than the Hubble
radius.  The same conclusion was reached using approximate methods by
Sakagami$^{107)}$ in a models with self interactions.
\section{Discussion}
\par
Let us first summarize the main results of the analysis of density
fluctuations in inflationary cosmology:
\item{-} Quantum vacuum fluctuations in the de Sitter phase of an inflationary
Universe are the source of perturbations.
\item{-} The quantum perturbations decoherence on scales outside the Hubble
radius and can hence be treated classically.
\item{-} The classical evolution outside the Hubble radius produces a large
amplification of the perturbations.  In fact, unless the particle physics
model contains very small coupling constants, the predicted fluctuations are
in excess of those allowed by the bounds on cosmic microwave anisotropies.
\par
Inflationary Universe models generically produce a scale invariant
\hfill \break Harrison-Zel'dovich spectrum$^{108)}$
$$
{\delta M\over M} (k , t_f (k) ) = {\rm const.} \eqno\eq
$$
This can be understood heuristically as follows: since the de Sitter phase is
time translation invariant, the evolution of each perturbation is identical
between the time it is produced and $t_i (k)$, the time it crosses the Hubble
radius.  Hence
$$
{\delta M\over M} \, (k, t_i (k) ) = {\rm const.} \eqno\eq
$$
However, since outside the Hubble radius the amplitude of the gauge invariant
measure of perturbations changes by a $k$ independent factor, (4.44) follows
immediately.
\par
It is not hard to construct models which give a different spectrum.  All that
is required is a significant time evolution of $H$ during the period of
inflation.
\par
I have chosen to present the analysis of fluctuations in inflationary cosmology
in three separate steps in order to highlight to crucial physics issues.
Having done this, it is possible to step back and construct a unified
analysis$^{59,109)}$ in which expectation values of gauge invariant variables
are propagated from $t \ll t_i (k)$ to $t_f (k)$ in a consistent way, and in
which the final values of the expectation values of the quadratic operators
are used to construct $T^{cl}_{\mu \nu} (\undertext{x} , t)$.

\chapter{Topological Defects in Cosmology}
\section{Introduction}
\par
In the previous two section we have seen that symmetry breaking phase
transitions in unified field theories arising in particle physics
({e.g.,} Grand Unified Theories$^{110)}$ (GUT)) do not lead to
inflation.  In general, the coupling constants which arise in the
effective potential for the scalar field $\varphi$ driving the phase
transition are too large to generate a period of slow rolling which
last more than one Hubble time $H^{-1} (t)$.  Nevertheless, there are
interesting remnants of the phase transition: topological
defects$^{16)}$.
\par
Consider a single component real scalar field with a typical symmetry breaking
potential
$$
V (\varphi) = {1\over 4} \lambda (\varphi^2 - \eta^2)^2 \eqno\eq
$$
Unless $\lambda \ll 1$ there
will be no inflation.  The phase transition will take place on a short time
scale $\tau < H^{-1}$, and will lead to correlation regions of radius $\xi <
t$ inside of which $\varphi$ is approximately constant, but outside of which
$\varphi$ ranges randomly over the vacuum manifold, the set of values
of $\varphi$ which minimizes $V(\varphi)$ -- in our example $\varphi
= \pm \eta$.  The correlation regions are separated by domain walls, regions in
space where $\varphi$ leaves the vacuum manifold ${\cal M}$ and where
therefore potential energy is localized.  Via the usual gravitational
force, this energy density can act as a
seed for structure.
\par
There are various types of local and global topological defects$^{16)}$
(regions of trapped energy density) depending on the number of components of
$\varphi$.  The more rigorous mathematical definition refers to the homotopy
of ${\cal M}$.  The words ``local" and ``global" refer to whether the symmetry
which is broken is a gauge or global symmetry.  In the case of local
symmetries, the topological defects have a well defined core outside of which
$\varphi$ contains no energy density in spite of nonvanishing gradients
$\nabla \varphi$:  the gauge fields $A_\mu$ can absorb the gradient,
{i.e.,} $D_\mu \varphi = 0$ when $\partial_\mu \varphi \neq 0$,
where the covariant derivative $D_\mu$ is defined by
$$
D_\mu = \partial_\mu - ig \, A_\mu \, , \eqno\eq
$$
$g$ being the gauge coupling constant.
Global topological defects, however, have long range density fields and
forces.
\endpage
\par
Table 1 contains a list of topological defects with their topological
characteristic.  A ``v" markes acceptable theories, a ``x" theories which are
in conflict with observations (for $\eta \sim 10^{16}$ GeV).
\vskip.5cm
\vskip4.5cm
\vskip.5cm
\par
Theories with domain walls are ruled out since a single domain wall
stretching
across the Universe today would overclose the Universe$^{111)}$.  Local
monopoles are
also ruled out since they would overclose the Universe$^{112)}$.  Local
textures are ineffective at producing structures (see Chapter 6).
\par
Let us demonstrate explicitly why stable domain walls are a
cosmological disaster.  If domain walls form during a phase transition
in the early Universe, it follows by causality (see however the caveats
of Refs. 113 and 114) that even today there will be at least one wall
per Hubble volume.  Assuming one wall per Hubble volume, the energy
density $\rho_{DW}$ of matter in domain walls is
$$
\rho_{DW} (t) \sim \eta^3 t^{-1} \, , \eqno\eq
$$
whereas the critical density $\rho_c$ is
$$
\rho_c = H^2 \, {3\over{8 \pi G}} \sim m^2_{p\ell} \, t^{-2} \, .
\eqno\eq
$$
Hence, for $\eta \sim 10^{16}$ GeV the ratio of (5.3) and (5.4) is
$$
{\rho_{DW}\over \rho_c} \, (t) \sim \, \left({\eta\over{m_{p\ell}}}
\right)^2 \, (\eta t) \sim 10^{52} \, . \eqno\eq
$$

The above argument depends in an essential way on the dimension of the
defect.  One cosmic string per Hubble volume leads to an energy
density $\rho_{cs}$ in string
$$
\rho_{cs} \sim \eta^2 \, t^{-2} \, . \eqno\eq
$$
Later in this section we shall see that the scaling (5.6) holds in the
cosmic string model.  Hence, cosmic strings do not lead to
cosmological problems.  In the contrary, since for GUT models with
$\eta \sim 10^{16}$ GeV
$$
{\rho_{cs}\over \rho_c} \sim \, \left({\eta\over m_{p \ell}} \right)^2
\sim 10^{-6} \, , \eqno\eq
$$
cosmic strings in these models could provide the seed perturbations
responsible for structure formation.

Theories with local monopoles are ruled out on cosmological
grounds$^{112)}$ (see again the caveats of Refs. 113 and 114) for
rather different reasons.  Since there are no long range forces
between local monopoles, their number density in comoving coordinates
does not decrease.  Since their contribution to the energy density
scales as $a^{-3} (t)$, they will come to dominate the mass of the
Universe, provided $\eta$ is sufficiently large.

Theories with global monopoles$^{115,116)}$ are not ruled out, since
there are long range forces between monopoles which lead to a
``scaling solution" with a fixed number of monopoles per Hubble
volume.

\section{Cosmic Strings}
\par
Consider a theory in which matter consists of a gauge field $A_\mu$ and a
complex scalar field $\phi$ whose dynamics is given by the Lagrangean
$$
{\cal L} = {1\over 2} \, D_\mu \phi D^\mu \phi - V (\phi) + {1\over 4} \,
F_{\mu \nu} \, F^{\mu \nu}  \eqno\eq
$$
where $F_{\mu \nu}$ is the field strength tensor.
The potential $V(\phi)$ has the symmetry breaking ``Mexican hat" shape (see
Figure 18):
$$
V (\phi) = {1\over 4} \lambda (|\phi|^2 - \eta^2)^2 \, . \eqno\eq
$$
Hence, the vacuum manifold ${\cal M}$, the space of minimum energy density
configurations, is a circle $S^1$.
\midinsert \vskip 6.5cm
\hsize=6in \raggedbottom
\noindent{\bf Figure 18.} The zero temperature potential energy of the
complex scalar field used in the cosmic string model.
\endinsert
\par
The theory described by (5.8) and (5.9) admits one dimensional topological
defects, cosmic strings.  In the Abelian Higgs model of this example
the string solutions were first found by Nielsen and Olesen$^{117)}$.
It is possible to construct string
configurations which are translationally invariant along the $z$ axis.  On a
circle $C$ in the $x-y$ plane with radius $r$, the boundary conditions
for $\phi$ are
$$
\phi (r, \theta) = \eta \, e^{i \theta} \, \eqno\eq
$$
where $\theta$ is the polar angle along $C$. The
configuration (5.10) has winding number 1: at all points of the circle, $\phi$
takes on values in ${\cal M}$, and as $\varphi$ varies from 0 to $2 \pi$,
$\phi$ winds once round ${\cal M}$.  By continuity it follows that there must
be a point $p$ on the disk $D$ bounded by $C$ where $\phi = 0$.  By
translational symmetry
there is a line of points with $\phi = 0$.  This line is the center of the
cosmic string.  The cosmic string is a line of trapped potential energy.  In
order to minimize the total energy given the prescribed topology
(i.e.,
winding number), the thickness of the string (i.e., radius over which $V
(\phi)$
deviates significantly from 0) must be finite.  As first shown in Ref.
117, the
width $w$ of a string is
$$
w \simeq \lambda^{-1/2} \eta^{-1} \, , \eqno\eq
$$
from which it follows that the mass per unit length $\mu$ is
$$
\mu \simeq \eta^2 \, , \eqno\eq
$$
{i.e.,} indpendent of the self-coupling constant $\lambda$.
\par
Cosmic strings arise in any model in which the vacuum manifold satisfies the
topological criterion
$$
\Pi_1 ({\cal M}) \neq {\bf 1} \, , \eqno\eq
$$
$\Pi_1$ being the first homotopy group.  Any field configuration $\phi
(\undertext{x})$ is characterized by an integer $n$, the element of $\Pi_1
({\cal M})$ corresponding to $\phi (\undertext{x})$.
\par
A cosmic string is an example of a \undertext{topological defect}.  A
topological defect has a well-defined core, a region in space where $\phi
\not\in {\cal M}$ and hence $V (\phi) > 0$.  There is an associated
winding number, and it is quantized.  Hence, a topological defect is stable.
Furthermore, topological defects exist for theories with global and local
symmetry groups.
\par
Cosmic strings are not the only topological defects.  In theories for which
the vacuum manifold ${\cal M}$ obeys $\Pi_0 ({\cal M}) \neq {\bf 1}$, two
dimensional defects -- domain walls -- exist.  An example is the theory of a
single real scalar field with symmetry breaking potential (5.9).

If the
theory contains three real scalar fields $\phi_i$ with potential (3.2) (if
$|\phi|^2 = \sum\limits^3_{i=1} \, \phi^2_i$), then $\Pi_2 ({\cal M})
\neq
{\bf 1}$ and monopoles result.  The construction of a monopole
configuration is illustrated in Fig. 19.
As the origin in physical space we select a point which is to become
the center of the monopole.  Consider a sphere $S_r$ of radius $r$
surrounding this point.  A spherically symmetric monopole
configuration is obtained by the identity map
$$
\eqalign{
& S_r \rightarrow {\cal M} = S^2 \cr
& (r, \, \theta, \, \varphi) \, \mathop \rightarrow\limits_{\varphi}
\, (\theta, \, \varphi) \, .} \eqno\eq
$$
This configuration has winding number 1.  Since the winding number of
maps $S^2 \rightarrow S^2$ is quantized, it cannot change as $r$
varies.  Thus, the only way to obtain a single valued field
configuration at $r = 0$ is for $\varphi (r, \, \theta, \, \varphi)$
to leave ${\cal M}$ as $r \rightarrow 0$.  In particular, there is a
point ({e.g.,} $r = 0$) for which $\varphi = 0$.  This is the
center of the monopole.  We see that monopoles are topological
defects: they contain a core, have quantized winding number and are
stable.
\midinsert \vskip 8.5cm
\hsize=6in \raggedright
\noindent {\bf Figure 19.} Construction of a monopole: left is
physical space, right the vacuum manifold.  The field configuration
$\phi$ maps spheres in space onto ${\cal M}$.  However, a core region
of space near the origin is mapped onto field values not in ${\cal
M}$.
\endinsert

\section{Global Textures}

Next, consider a theory of four real scalar fields given by the Lagrangean
$$
{\cal L} = {1\over 2} \partial_\mu \phi \partial^\mu \phi - V (\phi) \eqno\eq
$$
with
$$
V (\phi) = {1\over 4} \lambda \, \left( \sum\limits^4_{i=1} \, \phi^2_i -
\eta^2 \right)^2 \, . \eqno\eq
$$
In this case, the vacuum manifold is ${\cal M} = S^3$ with topology
$$
\Pi_3 ({\cal M}) \neq {\bf 1} \, , \eqno\eq
$$
and the corresponding defects are the global textures$^{16,118,119)}$.
\midinsert \vskip 10cm
\hsize=6in \raggedbottom
\noindent{\bf Figure 20.} Construction of a global texture: left is
physical space, right the vacuum manifold. The field configuration
$\phi$ is a map from space to the vacuum manifold (see text).
\endinsert
\par
Textures, however, are quite different than the previous topological defects.
The texture construction will render this manifest (Fig. 20).  To construct a
radially symmetric texture, we give a field configuration $\phi (x)$ which
maps physical space onto ${\cal M}$.  The origin 0 in space (an arbitrary point
will be the center of the texture) is mapped onto the north pole $N$ of ${\cal
M}$.  Spheres surrounding 0 are mapped onto spheres surrounding $N$.  In
particular, some sphere with radius $r_c (t)$ is mapped onto the equator
sphere of ${\cal M}$.  The distance $r_c (t)$ can be defined as the radius of
the texture.  Inside this sphere, $\phi (x)$ covers half the vacuum manifold.
Finally, the sphere at infinity is mapped onto the south pole of ${\cal M}$.
The configuration $\phi (\undertext{x})$ can be parameterized
by$^{119)}$
$$
\phi (x,y,z) = \left(\cos \chi (r), \> \sin \chi (r) {x\over r}, \>
\sin \chi (r) {y\over r}, \> \sin \chi (r) {z\over r} \right) \eqno\eq
$$
in terms of a function $\chi (r)$ with $\chi (0) = 0$ and $\chi (\infty) =
\pi$.  Note that at all points in space, $\phi (\undertext{x})$ lies in ${\cal
M}$.  There is no defect core.  All the energy is spatial gradient (and
possibly kinetic) energy.
\par
In a cosmological context, there is infinite energy available in an infinite
space.  Hence, it is not necessary that $\chi (r) \rightarrow \pi$ as $r
\rightarrow \infty$.  We can have
$$
\chi (r) \rightarrow \chi_{\rm max} < \pi \>\> {\rm as} \>\> r \rightarrow
\infty \, . \eqno\eq
$$
In this case, only a fraction
$$
n = {\chi_{\rm max}\over \pi} - {\sin 2 \chi_{\rm max}\over{2 \pi}} \eqno\eq
$$
of the vacuum manifold is covered:  the winding number $n$ is not quantized.
This is a reflection of the fact that whereas topologically nontrivial maps
from $S^3$ to $S^3$ exist, all maps from $R^3$ to $S^3$ can be deformed to
the trivial map.
\par
Textures in $R^3$ are unstable.  For the configuration described above, the
instability means that $r_c (t) \rightarrow 0$ as $t$ increases: the texture
collapses.  When $r_c (t)$ is microscopical, there will be sufficient energy
inside the core to cause $\phi (0)$ to leave ${\cal M}$, pass through 0 and
equilibrate at $\chi (0) = \pi$: the texture unwinds.
\par
A further difference compared to topological defects: textures are relevant
only for theories with global symmetry.  Since all the energy is in spatial
gradients, for a local theory the gauge fields can reorient themselves such as
to cancel the energy:
$$
D_\mu \phi = 0 \, . \eqno\eq
$$
\par
Therefore, it is reasonable to regard textures as an example of a new class of
defects, \undertext{semitopological defects}.  In contrast to topological
defects, there is no core, and $\phi (\undertext{x}) \epsilon {\cal M}$ for all
$\undertext{x}$.  In particular, there is no potential energy.  Second, the
winding number is not quantized, and hence the defects are unstable.  Finally,
they exist only in theories with a global interval symmetry.

\section{Kibble Mechanism}
\par
The Kibble mechanism$^{16)}$ ensures that in theories which admit
topological or semitopological defects, such defects will be produced
during a phase transition in the very early Universe.

Consider a mechanical toy model, first introduced by Mazenko, Unruh
and Wald$^{49)}$ in the context of inflationary Universe models, which
is useful in understanding the scalar field evolution.  Consider (see
Fig. 21) a lattice of points on a flat table.  At each point, a pencil
is pivoted.  It is free to rotate and oscillate.  The tips of nearest
neighbor pencils are connected with springs (to mimic the spatial
gradient terms in the scalar field Lagrangian).  Newtonian gravity
creates a potential energy $V(\varphi)$ for each pencil ($\varphi$ is
the angle relative to the vertical direction).  $V(\varphi)$ is
minimized for $| \varphi | = \eta$ (in our toy model $\eta = \pi /
2$).  Hence, the Lagrangian of this pencil model is analogous to that
of a scalar field with symmetry breaking potential (5.1).

\midinsert \vskip 5.5cm
\hsize=6in \raggedbottom
\noindent{\bf Figure 21.} The pencil model: the potential energy of a
simple pencil has the same form as that of scalar fields used for
spontaneous symmetry breaking.  The springs connecting nearest
neighbor pencils give rise to contributions to the energy which mimic
spatial gradient terms in field theory.
\endinsert

At high temperatures $T \gg T_c$, all pencils undergo large amplitude
high frequency oscillations.  However, by causality, the phases of
oscillation of pencils with large separation $s$ are uncorrelated.
For a system in thermal equilibrium, the length $s$ beyond which
phases are random is the correlation length $\xi (t)$.  However, since
the system is quenched rapidly, there is an opinion causality bound in
$\xi$:
$$
\xi (t) < t \, , \eqno\eq
$$
where $t$ is the causal horizon.

The critical temperature $T_c$ is the temperature at which the
thermal energy is equal to the energy a pencil needs to jump from
horizontal to vertical position.  For $T < T_c$, all pencils want to
lie flat on the table.  However, their orientations are random beyond
a distance of $\xi (t)$.  Below the Ginsburg temperature $T_G$, there
is insufficient thermal energy to excite a correlation volume into the
state $\varphi = 0$.  Domains of size$^{16,120)}$
$$
\xi (t_G) \sim \lambda^{-1} \eta^{-1} \eqno\eq
$$
freeze out.  The boundaries between these domains become topological
defects.

We conclude that in a theory in which a symmetry breaking phase
transitions satisfies the topological criteria for the existence of a
fixed type of defect, a network of such defects will form during the
phase transition and will freeze out at the Ginsburg temperature.  The
correlation length is initially given by (5.23), if the field
$\varphi$ is in thermal equilibrium before the transition.
Independent of this last assumption, the causality bound implies that
$\xi (t_G) < t_G$.

For times $t > t_G$ the evolution of the network of defects may be
complicated (as for cosmic strings) or trivial (as for textures).  In
any case (see the caveats of Refs. 113 and 114), the causality bound
persists at late times and states that even at late times, the mean
separation and length scale of defects is bounded by $\xi (t) \leq t$.

Applied to cosmic strings, the Kibble mechanism implies that at the
time of the phase transition, a network of cosmic strings with typical
step length $\xi (t_G)$ will form.  According to numerical
simulations$^{121)}$, about 80\% of the initial energy is in infinite
strings and 20\% in closed loops.

Note that the Kibble mechanism was discussed above in the context of a
global symmetry breaking scenario.  As pointed out in Ref. 122, the
situations is more complicated in local theories in which gauge field
can calcel spatial gradients in $\varphi$ in the energy functional,
and in which spatial gradients in $\varphi$ can be gauged away.
Nevertheless, as demonstrated numerically (in $2 + 1$ dimensions) in
Ref. 123 and shown analytically in Ref. 124, the Kibble mechanism also
applies to local symmetries.

The evolution of the cosmic string network for $t > t_G$ is
complicated (see Section 5.5).  The key processes are loop production
by intersections of infinite strings (see Fig. 22) and loop shrinking
by gravitational radiation.  These two processes combine to create a
mechanism by which the infinite string network loses energy (and
length as measured in comoving coordinates).  It will be shown that as
a consequence, the correlation length of the string network is always
proportional to its causality limit
$$
\xi (t) \sim t \, . \eqno\eq
$$
Hence, the energy density $\rho_\infty (t)$ in long strings is a fixed
fraction of the background energy density $\rho_c (t)$
$$
\rho_\infty (t) \sim \mu \xi (t)^{-2} \sim \mu t^{-2} \eqno\eq
$$
or
$$
{\rho_\infty (t)\over{\rho_c (t)}} \sim G \mu \, . \eqno\eq
$$

We conclude that the cosmic string network approaches a ``scaling
solution"$^{125,126)}$ in which the statistical properties of the
network are time independent if all distances are scaled to the
horizon distance.

Applied to textures$^{119)}$, the Kibble mechanism implies that on all
scales $r \geq t_G$, field configurations with winding number $n_W
\geq n_{cr}$ are frozen in with probability $p (n_{cr})$ per volume
$r^3$.  The critical winding number $n_{cr}$ is defined as the winding
number above which field configurations collapse and below which they
expand.  Only collapsing configurations form clumps of energy which
can accrete matter.

The critical winding $n_{cr}$ was determined numerically in Refs. 127
\& 128 and analytically in Ref. 129 (see also Ref. 130).  It is
slightly larger than 0.5.  The probability $p (n_{cr})$ can be
determined using combinational arguments$^{131)}$.

For $t > t_G$, any configuration on scale $\sim t$ with winding number
$n_W \ge n_{cr}$ begins to collapse (before $t$, the Hubble damping
term dominates over the spatial gradient forces, and the field
configuration is frozen in comoving coordinates).  After unwinding,
$\varphi (\undertext{x})$ is homogeneous inside the horizon.

The texture model thus also leads to a scaling solution: at all times
$t > t_G$ there is the same probability that a texture configuration
of scale $t$ will enter the horizon, become dynamical and collapse
with a typical time scale $t$.

\section{Evolution of the Cosmic String Network}
\par
If the evolution of the cosmic string network were trivial in the sense that
all strings would only stretch as the universe expands, there would be an
immediate cosmological disaster.  Consider a fixed comoving volume $V$ with a
string passing through.  The energy in radiation decreases as $a^{-1} (t)$
while the energy in string increases as $a (t)$.  Hence trivial evolution
would immediately lead to a string dominated universe, a cosmological
disaster.
\par
The equations of motion of a string are determined by the Nambu action
$$
S = - \mu \int d \sigma d \tau \left(- \det g^{(2)}_{ab} \right)^{1/2} \> \>
a, b = 0, 1\eqno\eq
$$
where $g^{(2)}_{ab}$ is the world sheet metric and $\sigma$ and
$\tau$ are the world sheet coordinates.  In flat space-time, $\tau$
can be taken to be coordinate time, and $\sigma$ is an affine
parameter along the string.  In terms of the string
coordinates $X^\mu (\sigma, \tau)$ and the metric $g^{(4)}_{\mu\nu}$ of
the background space-time,
$$
g^{(2)}_{ab} = X^\mu_{,a} X^v_{,b} g^{(4)}_{\mu\nu} \, . \eqno\eq
$$
{}From
general symmetry considerations,  it is possible to argue that the
Nambu action is the correct action.  However, I shall follow
Foerster$^{132)}$ and
Turok$^{133)}$ and give a direct heuristic derivation.  We start from a
general quantum field theory Lagrangian ${\cal L}_{QFT}$.  The action is
$$
S = \int d^4 y {\cal L}_{QFT} \left(\phi (y)\right)\eqno\eq
$$
We now assume the existence of a linear topological defect at $X^\mu (\sigma,
\tau)$.  The idea now is to change variables so that $\sigma$ and $\tau$ are
two of the new coordinates, and to expand $S$ to lowest order in $w/R$, where
$w$ is the width of the string and $R$ its curvature radius.  As the other new
coordinates we take coordinates $\rho^2$ and $\rho^3$ in the normal plane to
$X^\mu (\sigma, \tau)$.  Thus the coordinate transformation takes the
coordinates $y^\mu (\mu = 0, 1, 2, 3)$ to new ones $\sigma^3 = (\tau, \sigma,
\rho^2, \rho^a)$:
$$
y^\mu (\sigma^a) = X^\mu (\sigma, \tau) + \rho^i n^\mu_i (\sigma,
\tau)\eqno\eq
$$
where $i = 2,3$ and $n^\mu_i$ are the basis vectors in the normal plane to the
string world sheet.  The measure transforms as
$$
\int d^4 y = \int d \sigma d \tau d \rho^2 d \rho^3 (\det M_a^\mu)\eqno\eq
$$
with
$$
M^\mu_a = \, {\partial y^\mu\over{\partial \sigma^a}} = \, \pmatrix{\partial
X^\mu/\partial (\sigma, \tau)\cr
n^\mu_i\cr} + O (\rho)\, .\eqno\eq
$$
The determinant can easily be evaluated using the following trick
$$
\det M^\mu_a = \, \left( - \det \eta_{\mu \nu} M^\nu_a M^\nu_b \right)^{1/2}
\equiv \sqrt{- \det D_{ab}}\eqno\eq
$$
$$
D = \, \pmatrix{{\partial x^\mu\over{\partial (\sigma, \tau)}} \, {\partial
X^\nu\over{\partial (\sigma, \tau)}} \eta_{\mu \nu} & {\partial X^\mu\over
{\partial (\sigma , \tau)}} n^\nu_b \eta_{\mu\nu}\cr
{\partial X^\mu\over{\partial (\sigma, \tau)}} n^\nu_a \eta_{\mu\nu} & n^\mu_a
n^\nu_b \eta_{\mu\nu}\cr} = \pmatrix{X^\mu_{,a} X^\nu_{,b} \eta_{\mu\nu} & 0\cr
0 & \delta_{ab}\cr} + 0 \, \left({w\over R}\right) \eqno\eq
$$
Hence
$$
\eqalign{S &= \int d \sigma d \tau \left( - \det g^{(2)}_{ab} \right)^{1/2}
\int d \rho^2 d \rho^3 {\cal L} (y (\sigma, \tau, \rho^2, \rho^3)) + O
\left({w\over R} \right)\cr
&= - \mu \int d \sigma d \tau \left( - \det g^{(2)}_{ab} \right)^{1/2} + O
\left({w\over R}\right)\, . }\eqno\eq
$$
$- \mu$ is the integral of {\cal L} in the normal plane of $X$.  To first
order in $w/R$, it equals the integral of $-{\cal H}$; hence it is the mass per
unit length.
\par
This derivation of the Nambu action is instructive as it indicates a method
for calculating corrections to the equations of motion of the string when
extra fields are present, \eg\ for superconducting cosmic strings.  It also
gives a way of calculating the finite thickness corrections to the equations
of motion which will be important at cusps (see below).
\par
In flat space-time we can consistently choose $\tau = t, \dot x \cdot x^\prime
= 0$ and $\dot x^2 + x^{\prime^2} = 0$.  The equations of motion derived from
the Nambu action then become
$$
\ddot {\undertext{x}} - \undertext{x}^{\prime\prime} = 0\, . \eqno\eq
$$
where $\prime$ indicates the derivative with respect to $\sigma$.  The general
solution can be decomposed into a left moving and a right moving mode
$$
\undertext{x} (t, \sigma) = {1\over 2} \, \left[ \undertext{a} (\sigma - t) +
\undertext{b} (\sigma + t ) \right] \eqno\eq
$$
The gauge conditions imply
$$
\dot {\undertext{a}}^2 = \dot {\undertext{b}}^2 = 1\eqno\eq
$$
For a loop, $\undertext{x} (\sigma, t)$ is periodic and hence the time average
of $\dot {\undertext{a}}$ and $\dot {\undertext{b}}$ vanish.  $\dot
{\undertext{a}}$
and $\dot {\undertext{b}}$ are closed curves on the unit sphere with vanishing
average.  Two such curves generically intersect if they are
continuous$^{134)}$.  An intersection corresponds to a point with
$\undertext{x}^\prime = 0$ and $\dot {\undertext{x}} = 1$.  Such a point moving
at the speed of light is called a cusp.  $\dot {\undertext{x}} (\sigma, t)$
need
not be continuous.  Points of discontinuity are called kinks.  Note that both
cusps and kinks will be smoothed out by finite thickness
effects$^{135)}$.
\par
The Nambu action does not describe what happens when two strings hit.  This
process has been studied numerically for both global$^{136)}$ and
local$^{137)}$ strings.  The authors set up scalar field configurations
corresponding to two strings approaching one another and evolve the complete
classical scalar field equations.  The result of the analysis is that strings
do not cross but exchange ends, provided the relative velocity is smaller than
0.9.  Thus, by self intersecting, an infinite string will split off a loop
(Figure 22).  An important open problem is to understand this process
analytically.  For a special value of the coupling constant Ruback has
recently given a mathematical explanation$^{138)}$ (see also Shellard and
Ruback in Ref. 137).
\midinsert \vskip 3.5cm
\hsize=6in \raggedright
\noindent {\bf Figure 22.}  Formation of loops by self intersection of
infinite strings.  According to the original cosmic string scenario, loops
form with radius $R$ determined by the instantaneous correlation length of the
infinite string network.
\endinsert
\par
There are two parts to the nontrivial evolution of the cosmic string network.
Firstly, loops are produced by self intersections of infinite strings.  Loops
oscillate due to the tension and slowly decay by emitting gravitational
radiation$^{139)}$.  Combining the two steps we have a process by which energy
is transferred from the cosmic string network to radiation.
\par
There are analytical indications that a stable ``scaling solution"
(already described in Section 5.3) for the
cosmic string network exists$^{5)}$.  It is given by on the order 1 infinite
string segment crossing every Hubble volume.  The correlation length $\xi (t)$
of an infinite string is of the order $t$.  Hence, at time $t$ loops of radius
$R \sim t$ are produced, of the order 1 loop per Hubble volume per expansion
time.  A heuristic argument for the scaling solution is due to Vilenkin.  Take
$\tilde \nu (t)$ to be the mean number of infinite string segments per Hubble
volume.  Then the energy density in infinite strings is
$$
\rho_\infty (t) = \mu \tilde \nu (t) t^{-2} \eqno\eq
$$
The number of loops $n(t)$ produced per unit volume is given by
$$
{d n (t)\over{dt}} = c \tilde \nu^2 t^{-4} \eqno\eq
$$
where $c$ is a constant of the order $1$.  Conservation of energy in strings
gives
$$
{d \rho_\infty (t)\over{dt}} + {3\over{2 t}} \, \rho_\infty (t) = - c^\prime
\mu t \, {dn\over{dt}} = - c^\prime \mu \tilde \nu^2 t^{-3} \eqno\eq
$$
or, written as an equation for $\tilde \nu (t)$
$$
\tilde {\dot \nu} - \, {\tilde \nu\over{2 t}} = - cc^\prime \tilde \nu^2 t^{-
1}\eqno\eq
$$
Thus if $\tilde \nu \gg 1$ then $\tilde {\dot \nu} < 0$ while if $\tilde \nu
\ll 1$ then $\tilde {\dot \nu} > 0$.  Hence there will be a stable solution
with $\tilde \nu \sim 1$.

The precise value of $\tilde \nu$ must be determined in numerical simulations.
These simulations are rather difficult because of the large dynamic range
required and due to singularities which arise in the evolution equations near
cusps.  In the radiation dominated epoch, $\tilde \nu$ is still uncertain by a
factor of about 10.  The first results were reported in Ref. 140.
More recent results are due three groups.
Bennett and Bouchet$^{141)}$ and Allen and Shellard$^{142)}$ are
converging on
a value $10 < \tilde \nu < 20$, whereas Albrecht and Turok$^{143)}$ obtain a
value which is about 100.
\section{Scaling Solution for Strings}
\par
The scaling solution for the infinite strings implies that the network of
strings looks the same at all times when scaled to the Hubble radius.  This
should also imply that the distribution of cosmic string loops is scale
invariant in the same sense.  At present, however, there is no convincing
evidence from numerical simulations that this is really the case.
\par
A scaling solution for loops implies that the distribution of $R_i (t)$, the
radius of loops at the time of formation, is time independent after dividing
by $t$.  To simplify the discussion, I shall assume that the
distribution in monochromatic, \ie\
$$
R_i (t)/t = \alpha\, . \eqno\eq
$$
 From Figure 22, we expect $\alpha \sim 1$.  The numerical simulations,
however, now give $\alpha < 10^{-2}$~$^{141,142)}$.
\par
 From the scaling solution (5.39) for the infinite strings we can derive the
scaling solution for loops.  We assume that the energy density in long strings
-- inasmuch as it is not redshifted -- must go into loops.  $\beta$ shall be a
measure for the mean length $\ell$ in a loop of ``radius" $R$
$$
\ell = \beta R\, . \eqno\eq
$$
Since per expansion time and Hubble volume about 1 loop of radius $R_i (t)$ is
produced, we know that the number density in physical coordinates of loops of
radius $R_i (t)$ is
$$
n (R_i (t), t) = ct^{-4}\eqno\eq
$$
with a constant $c$ which can be calculated from (5.39), (5.43) and
(5.44).
Neglecting gravitational radiation, this number density simply redshifts
$$
n (R,t) = \, \left({z (t)\over{z (t_f (R))}} \right)^3 n (R, t_f (R))\, ,
\eqno\eq
$$
where $t_f (R)$ is the time when loops of radius $R$ are formed.  Isolating
the $R$ dependence, we obtain
$$
n (R, t) \sim R^{-4} z (R)^{-3}\eqno\eq
$$
where $z (R)$ is the redshift at time $t=R$.  We have the following special
cases:
$$
\eqalign{n (R, t) \sim R^{-5/2} t^{-3/2} \qquad & t < t_{eq}\cr
n (R, t) \sim R^{-5/2} t_{eq}^{1/2} t^{-2} \qquad & t > t_{eq} \, , \, t_f
(R) < t_{eq}\cr
n (R, t) \sim R^{-2} t^{-2} \qquad & t > t_{eq} \, , \, t_f (R) > t_{eq} \,
.}\eqno\eq
$$
\par
The proportionality constant $c$ is
$$
c = {1\over 2} \beta^{-1} \alpha^{-2} \tilde \nu\eqno\eq
$$
(see \eg\ Ref. 144).  In deriving (5.49) it is important to note that $n (R_i
(t), t) dR_i$ is the number density of loops in the radius interval $[R_i ,
R_i + dR_i]$.  Hence, in the radiation dominated epoch
$$
n (R, t) = \nu R^{- 5/2} t^{-3/2} \eqno\eq
$$
with
$$
\nu = {1\over 2} \beta^{-1} \alpha^{1/2} \tilde \nu \, .\eqno\eq
$$
\par
 From (5.51) we can read off the uncertainties in $\nu$ based on the
uncertainties in the numerical results.  Both $\alpha^{1/2}$ and $\tilde \nu$
are determined only up to one order of magnitude.  Hence, any quantitative
results which depend on the exact value of $\nu$ must be treated with a grain
of salt.
\par
Graviational radiation leads to a lower cutoff in $n (R, t)$.  Loops with
radius smaller than this cutoff were all formed at essentially the same time
and hence have the same number density.  Thus, $n (R)$ becomes flat.  The
power in gravitational radiation $P_G$ can be estimated using the quadrapole
formula$^{145)}$.  For a loop of radius $R$ and mass $M$
$$
P_G = {1\over 5} G < \dot{\ddot Q} \dot{\ddot Q} > \, , \eqno\eq
$$
where $Q$ is the quadrapole moment, $Q \sim MR^2$, and since the frequency of
oscillation is $\omega = R^{-1}$
$$
P_G \sim G (MR^2)^2 \omega^6 \sim (G \mu) \mu \, . \eqno\eq
$$
\par
Even though the quadrapole approximation breaks down since the loops move
relativistically, (5.53) gives a good order of magnitude of the power of
gravitational radiations.  Improved calculations$^{139)}$ give
$$
P_G = \gamma (G \mu) \mu\eqno\eq
$$
with $\gamma \sim 50$.  (5.44) and (5.54) imply that
$$
\dot R = \tilde \gamma G \mu\eqno\eq
$$
with $\tilde \gamma \equiv \gamma/\beta \sim 5$ (using $\beta \simeq
10$~$^{140)}$).  Note that the rate of decrease is constant.  Hence,
$$
R (t) = R_i - (t - t_i) \tilde \gamma G \mu\eqno\eq
$$
and the cutoff loop radius is
$$
R_c \sim \tilde \gamma G \mu t_i\, . \eqno\eq
$$
\par
Let us briefly summarize the scaling solution
\item{1)} At all times the network of infinite strings looks the same when
scaled by the Hubble radius.  A small number of infinite string segments cross
each Hubble volume and $\rho_\infty (t)$ is given by (5.39).
\item{2)}  There is a distribution of loops of all sizes $0 \le R < t$.
Assuming scaling for loops, then
$$
n (R, t) = \nu R^{-4} \, \left({z (t)\over{z (R)}}\right)^3 \, , \> R
\> \epsilon \> [ \tilde \gamma G \mu t,  \alpha t]\eqno\eq
$$
where $\alpha^{-1} R$ is the time of formation of a loop of radius $R$.  Also
$$
n (R, t) = n ( \tilde \gamma G \mu t,  t) \, , \> R < \tilde \gamma G \mu t\,
. \eqno\eq
$$

Although the qualitative characteristics of the cosmic string scaling
solution are well established, the quantitative details are not.  The
main reason for this is the fact that the Nambu action breaks down at
kinks and cusps.  However, kinks and cusps inevitably form and are
responsible for the small scale structure on strings.  In fact, coarse
graining by integrating out the small scale structure may give an
equation of state for strings which deviates from that of a Nambu
string$^{146)}$.  Attempts at understanding the small scale structure
on strings are at present under way$^{147)}$.

\chapter{Topological Defect Models of Structure Formation}
\section{Power Spectrum}
\par
To begin this section we shall demonstrate that the scaling solution
for topological defects discussed in the previous section leads to a
scale-invariant spectrum of perturbations.

The most commonly used function describing the ensemble of
perturbations is the spectrum $P(k)$.  By definition, the power
spectrum is the square of the modulus of the Fourier space density
contrast
$$
P (k) = \big| {{\delta \rho} \over \rho} (k) \big|^2 \, . \eqno\eq
$$
Hence (from (4.4)), $P(k)$ is related to the r.m.s. mass fluctuations
$(\delta M/M)\, (k,t)$ on physical length scale
$$
\lambda_k = a (t) \, {2\pi\over k} \eqno\eq
$$
at time $t$ via
$$
\left( {\delta M\over M} \right)^2 \, (k,t) \simeq k^3 P (k) \, .
\eqno\eq
$$

The scaling solution for topological defect models implies that when
measured at the time $t_H (k)$ when the wavelength $\lambda_k$ equals
the Hubble radius, the r.m.s. mass perturbation $(\delta M/M) \, (k)$
is independent of $k$, i.e.,
$$
{\delta M\over M} \, (k, t_H (k)) = {\rm const} \, . \eqno\eq
$$
This is because at any time $t$, a constant fraction of the mass $M$
inside the Hubble radius is contained in the topological defects.  For
example, one cosmic string of length $t$ contains mass $\delta M= \mu
t$ compared to the total mass $M \sim t^3 \rho (t) \sim t$ inside the
Hubble radius, thus leaving the ratio $\delta M/M$ time independent.

Eq. (6.4) is the same result as is obtained for inflationary Universe
models.  Hence, we conclude that all three main models of structure
formation: adiabatic random phase perturbations from inflation, cosmic
strings, and global textures, to a first approximation produce a
scale-invariant spectrum.

To convert (6.4) into an expression for the power spectrum $P(k)$, we
use the fact that $\delta M/M$ grows as the scale factor $a(t)$ during
the matter dominated epoch on scales smaller than the Hubble
radius$^{30)}$
$$
{\delta M\over M} \, (k,t) = \left({t\over{t_H (k)}} \right)^{2/3} \,
{\delta M\over M} \, (k, t_H (k)) \, . \eqno\eq
$$
On scales larger than the Hubble radius at $t_{eq}$
$$
t_H (k) = 2 \pi k^{-1} a (t_H (k)) \sim t_H^{2/3} (k) \, k^{-1} \, ,
\eqno\eq
$$
and hence
$$
t_H (k) \sim k^{-3} \, . \eqno\eq
$$
Therefore, combining (6.4), (6.5) and (6.7)
$$
{\delta M\over M} \, (k,t) \sim k^2 \, . \eqno\eq
$$
 From (6.3) it follows that
$$
P (k) \sim k^n \eqno\eq
$$
with $n = 1$.

Recently, there has been some interest in deviations from scale
invariance.  In models of inflation, a deviation comes about$^{148)}$
because $H$ decreases slowly during inflation.  In topological defect
models, numerical$^{149,150)}$ and semi-analytical$^{151)}$ studies
have also shown small deviations from scale invariance.  These
deviations, however, all small and quite model dependent.

The power spectrum $P(k)$ can in principle be tested in maps of CMB
anisotropies.  Any cosmological model with density inhomogeneities
gives rise to temperature anisotropies via the Sachs-Wolfe$^{152)}$
effect (see Fig. 23).  Photons last scattered at temperature $T_{rec}$
were emitted at slightly different times in different directions, and
hence arrive at the earth having travelled not quite the same distance
and hence having direction dependent temperature.
\midinsert \vskip 5.5cm
\hsize=6in \raggedright
\noindent {\bf Figure 23.} The Sachs-Wolfe effect: density
perturbations at the time of last scattering imply that light rays
$\gamma$ from different directions in the sky have travelled a
different distance and hence arrive at the observer $\theta$ with
varying temperature.\endinsert

For adiabatic perturbations, the present temperature fluctuations on
angular scale $\theta$ are determined$^{30,15)}$ by the gravitational
potential at $t_{rec}$, which in turn is equal to the density
perturbation at $t_{eq}$:
$$
{\delta T\over T} (\theta) \simeq {1\over 3} \, {\delta \rho\over
\rho} \, (k (\theta), \, t_{eq} ) \, . \eqno\eq
$$

The recent discovery$^{12)}$ of anisotropies in the CMB can be used to
normalize the power spectrum.  The observational results are
$$
{\delta T\over T} \, (90^\circ) \simeq 5 \cdot 10^{-6} \eqno\eq
$$
for the quadrupole moment.  The beam width of the COBE satellite is
large, and hence only anisotropies for $\theta > 10^\circ$ can be
determined.  The analysis of the data gives agreement with a power
spectrum with
$$
n = 1.1 \pm 0.5 \, . \eqno\eq
$$

The results (6.11) and (6.12) cannot be used to differentiate between
the three theories discussed in these lectures, since the models all
predict a similar slope of $P(k)$.  However, the amplitude of $\delta
T/T$ can be used to normalize the power spectrum in any given model.
Large scale structure observations provide an independent
normalization.  These two normalizations must be consistent in order
for a theoretical model to work.  For cosmic strings, the
normalizations of $P(k)$ agree$^{149,151)}$ well, for an unbiased
texture model, the normalization factors differ by about 3$^{150)}$.

In the near future, maps of CMB anisotropies will be obtained$^{153)}$
which are signal dominated in every pixel (the COBE maps are dominated
by noise).  At that point, statistical$^{154,155)}$ measures of CMB
maps can be evaluated which can pick out the non-Gaussian
signatures of topological defect models.  Indeed, non-random phases of
the Fourier modes of $\delta \rho$ are the key feature of such models
(see Fig. 24).
\midinsert \vskip 4.8cm
\hsize=6in \raggedright
\noindent {\bf Figure 24.} A comparison of density perturbations along
a line in space between (a) random phase models and (b) theories based
on nonadiabatic seeds. \endinsert

\section{Large-Scale Structure Data}
\par
Here, I shall briefly mention some of the data obtained from optical
telescopes
which at the present time seems
most relevant to cosmology.
\par
First, however, it is important to note a discrepancy between theory and
observations.  Observations become increasingly uncertain on increasing length
scales, whereas theoretical predictions decrease in accuracy as the length
scale decreases, the reason being that on large scales linear theory is
applicable and gravity is the only force which needs to be considered, whereas
on smaller scales nonlinear and messy nongravitational effects become
important.  However, this state of affairs also implies that progress either
on the theoretical or observational front will yield double benefit.  For
example, new data on larger scales not only increases the amount of data
available, but also leads into a regime with smaller theoretical
uncertainties.  Since we know that observers will be providing lots of new
data in the next few years, we can be sure that cosmology will remain an
extremely exciting field.
\par
There is a lot of new data on the large-scale structure of the Universe.
Redshift surveys have provided three dimensional maps of the distribution of
galaxies.  An example is the recent Center for Astrophysics (CFA)
survey$^{11)}$ of slices on the northern celestial sphere which show
overdense sheets of galaxies with dimensions $(50 \times 50 \times
5)$Mpc$^3$ (with $h =1$) separated by large voids.  There is evidence for
superclusters$^{156)}$, filaments longer than $50h^{-1}$Mpc$^{157)}$, voids of
diameter $60h^{-1}$Mpc$^{158)}$ and overdense sheets of
galaxies$^{159)}$ in
regions of the sky different from that covered in the CFA survey.
\par
A second window on the large-scale structure of the Universe comes from
measuring the peculiar velocities of galaxies averaged over large
regions$^{160)}$ from which one can infer information about the magnitude of
density perturbations on large scales$^{161)}$.  Preliminary observations
indicate surprisingly large velocities on scales of $60h^{-1}$Mpc, although
consensus on the interpretation of the data is still
lacking$^{160,162)}$.
\par
On the scale of clusters the data is still rather uncertain.  The quantities
one would like to focus on is the overall mass scale (the mean mass of an
object which satisfies a fixed operational definition of a cluster), the
multiplicity function and the two point (and in the future higher point)
correlation functions.
\par
A cluster of galaxies (Abell$^{163)}$ cluster) is defined as a region in the
sky with more than 50 bright galaxies in a sphere of radius $1.5h^{-1}$Mpc.
(To be compared to the mean separation of bright galaxies$^{164)}$ which is
about $5h^{-1}$Mpc).  The mean separation of clusters is about $50 h^{-1}$Mpc,
and their masses are of the order
$10^{15}$M$\odot$~$^{165)}$.
\par
The multiplicity function $n(M)$ of clusters gives the number density of
clusters of mass $M$ per unit physical volume.  $n(M) dM$ is the number
density of objects in the mass interval $[M, M+dM]$.
\par
The two point correlation function $\xi (r)$ of clusters measures the
nonrandomness of the distribution of clusters and is defined by
$$
\xi (r) = \, < {n (r) - n_0\over{n_0}} > \, , \eqno\eq
$$
where $n(r)$ is the number density at a fixed distance $r$ from a given object
and $n_0$ is the average density, and pointed brackets indicate
averaging over the objects.
\par
The initial observational results $^{166)}$ (Figure 25) gave
$$
\xi (r) \simeq \left({r\over{r_0}}\right)^{-2}\eqno\eq
$$
with correlation length $r_0 \sim 25 h^{-1}$Mpc.  Recently$^{167)}$, some
criticism of these results has been raised.  However, it is unlikely that the
entire effect is fictitious.
\midinsert \vskip 10cm
\hsize=6in \raggedright
\noindent {\bf Figure 25.}  The observed cluster correlation function plotted
against relative separation
(from Ref. 166).  The solid circles are the data points.  Shown is the fit to
a $r^{-2}$ slope.
\endinsert
\midinsert \vskip 8cm
\hsize=6in \raggedright
\noindent {\bf Figure 26.}  The mass function of galaxies (determined from the
luminosity function assuming constant mass to light ratio) from Refs.
168 ((a)
from Bahcall, (b) from Binggeli).
\endinsert
\par
For galaxies, similar quantities can be measured and compared with
observations.  Large spiral galaxies have masses $M \sim 10^{12}M \odot$.
Their mean separation is about $5 h^{-1}$Mpc.  The mass
function$^{168)}$
(Figure 26) gives the distribution of masses, and the two point correlation
function describes the nonrandomness of the distribution.  The two
point correlation function of galaxies has the same power law form as
(6.14), with correlation length $r_0 \sim 8$Mpc.
For galaxies, one
can in addition probe the interior structure by measuring the velocity
rotation curve $v(r)$ which is related to the density profile (baryonic plus
dark matter) by
$$
v (r) \sim r^2 \rho (r)\, .\eqno\eq
$$
Observations$^{169)}$ indicate that far beyond the disk radius, $v (r) \sim
{\rm const}$ (Figure 27) which indicates the existence of dark matter with a
density profile
$$
\rho (r) \sim r^{-2}\, . \eqno\eq
$$
We can also measure the angular momentum of galaxies.  Typical numbers for
large spirals are in the range$^{170)}$ $10^{73} - 10^{75} cm^2 g\, s^{-
1}$.
\midinsert \vskip 5.5cm
\hsize=6in \raggedright
\noindent {\bf Figure 27.}  A typical velocity rotation curve (for NGC488 HI
data, taken from Ref. 169).  The radius is in $kpc$, the velocity in $kms^{-
1}$.
\endinsert

\section{Cosmic Strings and Structure Formation}

Starting point of the structure formation scenario in the cosmic
string theory is the scaling solution for the cosmic string network,
according to which at all times $t$ (in particular at $t_{eq}$, the
time when perturbations can start to grow) there will be a few long
strings crossing each Hubble volume, plus a distribution of loops of
radius $R \ll t$ (see Fig. 28).
\midinsert \vskip 6.5cm
\hsize=6in \raggedright
\noindent {\bf Figure 28.}  Sketch of the scaling solution for the
cosmic string network.  The box corresponds to one Hubble volume at
arbitrary time $t$. \endinsert

\par
The cosmic string model admits three mechanisms for structure
formation:  loops, filaments wakes.  Cosmic string loops have the same time
averaged
field as a point source with mass
$$
M (R) = \beta R \mu \, , \eqno\eq
$$
$R$ being the loop radius and $\beta \sim 2 \pi$.  Hence, loops will be seeds
for spherical accretion of dust and radiation.

For loops with $R \leq t_{eq}$, growth of perturbations in a model
dominated by cold dark matter starts at $t_{eq}$.  Hence, the mass at
the present time will be
$$
M (R, \, t_0) = z (t_{eq}) \beta \, R \mu \, . \eqno\eq
$$

In the original cosmic string model$^{125,126,171)}$ it was assumed
that loops dominate over wakes.  In this case, the theory could be
normalized (i.e., $\mu$ could be determined) by demanding that loops
with the mean separation of clusters $d_{cl}$ (from the discussion in
Section 5.6 it follows that the loop radius $R (d_{cl})$ is determined
by the mean separation) accrete the correct mass, i.e., that
$$
M (R (d_{cl}), t_0) = 10^{14} M_{\odot} \, . \eqno\eq
$$
This condition yields$^{171)}$
$$
\mu \simeq 10^{36} {\rm GeV}^2 \eqno\eq
$$
Thus, if cosmic strings are to be relevant for structure formation,
they must arise due to symmetry breaking at energy scale $\eta
\simeq 10^{16}$GeV.  This scale happens to be the scale of unification
of weak, strong and electromagnetic interactions.  It is tantalizing
to speculate that cosmology is telling us that there indeed was new
physics at the GUT scale.
\midinsert \vskip 3.5cm
\hsize=6in \raggedbottom
\noindent{\bf Figure 29.} Sketch of the mechanism by which a long
straight cosmic string moving with velocity $v$ in transverse
direction through a plasma induces a velocity perturbations $\Delta v$
towards the wake. Shown on the left is the deficit angle, in the
center is a sketch of the string moving in the plasma, and on the
right is the sketch of how the plasma moves in the frame in which the
string is at rest.
\endinsert
\par
The second mechanism involves long strings moving with relativistic
speed in their normal plane which give rise to
velocity perturbations in their wake.  The mechanism is illustrated in
Fig. 29:
space normal to the string is a cone with deficit angle$^{172)}$
$$
\alpha = 8 \pi G \mu \, . \eqno\eq
$$
If the string is moving with normal velocity $v$ through a bath of dark
matter, a velocity perturbation
$$
\delta v = 4 \pi G \mu v \gamma \eqno\eq
$$
[with $\gamma = (1 - v^2)^{-1/2}$] towards the plane behind the string
results$^{173)}$.  At times after $t_{eq}$, this induces planar overdensities,
the most
prominent (i.e. thickest at the present time) and numerous of which were
created at $t_{eq}$, the time of equal matter and
radiation$^{174,175)}$.  The
corresponding planar dimensions are (in comoving coordinates)
$$
t_{eq} z (t_{eq}) \times t_{eq} z (t_{eq}) v \sim (40 \times 40 v) \,
{\rm Mpc}^2
\, . \eqno\eq
$$

The thickness $d$ of these wakes can be calculated using the
Zel'dovich approximation$^{30)}$.  The result is
$$
d \simeq G \mu v \gamma (v) z (t_{eq})^2 \, t_{eq} \simeq 4 v \, {\rm
Mpc} \, . \eqno\eq
$$
\par
Wakes arise if there is little small scale structure on the string.
In this case, the string tension equals the mass density, the string
moves at relativistic speeds, and there is no local gravitational
attraction towards the string.

In contrast, if there is small scale structure on strings,
then$^{146)}$ the string tension $T$ is smaller than the mass per unit
length $\mu$ and the metric of a string in $z$ direction becomes
$$
ds^2 = (1 + h_{00}) (dt^2 - dz^2 - dr^2 - (1 - 8G \mu) r^2 dy^2 )
\eqno\eq
$$
with
$$
h_{00} = 4G (\mu - T) \ln \, {r\over r_0} \, , \eqno\eq
$$
$r_0$ being the string width.  Since $h_{00}$ does not vanish, there
is a gravitational force towards the string which gives rise to
cylindrical accretion, thus producing filaments.

As is evident from the last term in the metric (6.25), space
perpendicular to the string remains conical, with deficit angle given
by (6.21).  However, since the string is no longer relativistic, the
transverse velocities $v$ of the string network are expected to be
smaller, and hence the induced wakes will be shorter and thinner.

Which of the mechanisms -- filaments or wakes -- dominates is
determined by the competition between the velocity induced by $h_{00}$
and the velocity perturbation of the wake.  The total velocity
is$^{177)}$
$$
u = - {2 \pi G (\mu - T)\over{v \gamma (v)}} - 4 \pi G \mu v \gamma
(v) \, , \eqno\eq
$$
the first term giving filaments, the second producing wakes.  Hence,
for small $v$ the former will dominate, for large $v$ the latter.

By the same argument as for wakes, the most numerous and prominent
filaments will have the distinguished scale
$$
t_{eq} z (t_{eq}) \times d_f \times d_f \eqno\eq
$$
where $d_f$ can be calculated using the Zel'dovich approximation.

\section{Textures and Structure Formation}

Starting point of the texture scenario of structure formation is the
scaling solution for textures: at any time $t$, there is a fixed
probability $p$ that a texture configuration is entering the Hubble
radius and starting to collapse.

In the texture model it is the contraction of the field configuration which
leads to density perturbations.  At the time when the texture enters the
horizon, an isocurvature perturbation is established:  the energy density in
the scalar field is compensated by a deficit in radiation.  However, the
contraction of the scalar field configuration leads to a clumping of gradient
and kinetic energy at the center of the texture$^{178)}$ (Fig. 30).  This, in
turn, provides the
seed perturbations which cause dark matter and radiation to collapse in a
spherical manner$^{179,180)}$.

\midinsert \vskip 5.5cm
\hsize=6in \raggedbottom
\noindent {\bf Figure 30}: A sketch of the density perturbation produced
by a collapsing texture.  The left graph shows the time evolution of
the field $\chi (r)$ as a function of radius $r$ and time (see
(5.18)).  The contraction of $\chi (r)$ leads to a spatial gradient
energy perturbation at the center of the texture, as illustrated on
the right.  The energy is denoted by $\rho$.  Solid lines denote the
initial time, dashed lines are at time $t + \Delta t$, and dotted
lines correspond to time $t + 2 \Delta t$, where $\Delta t$ is a
fraction of the Hubble expansion time (the typical time scale for the
dynamics).
\endinsert

As in the cosmic string model, also in the global texture scenario the
length scale of the dominant structures is the comoving Hubble radius
at $t_{eq}$.  Textures generated at $t_{eq}$ are the most numerous,
and the perturbations induced by them have the most time to grow.

In both cosmic string and texture models, the fluctuations are non-Gaussian,
which means that the Fourier modes of the density perturbation $\delta \rho$
have nonrandom phases.  Most inflationary Universe models, in contrast,
predict (in linear theory) random phase fluctuations which can be
viewed as a superposition of small amplitude plane wave perturbations with
uncorrelated phases (for some subtle issues see Refs. 181 and 182).
\par
Before discussing some key observations which will allow us to distinguish
between the different models, I will discuss the role of dark matter.  The key
issue is free streaming.  Recall that cold dark matter consists of particles
which have negligible velocity $v$ at $t_{eq}$, the time when sub-horizon
scale perturbations can start growing:
$$
v (t_{eq}) \ll 1 \qquad {\rm (CDM)} \, . \eqno\eq
$$
For hot dark matter, on the other hand:
$$
v (t_{eq}) \sim 1 \qquad {\rm (HDM)} \, . \eqno\eq
$$
Due to their large thermal velocities, it is not possible to establish
HDM perturbations at early times on small scales.  Fluctuations are
erased by free streaming on all scales smaller than the free streaming
length
$$
\lambda_J^c (t) = v (t) z (t) t \eqno\eq
$$
(in comoving coordinates).  For $t > t_{eq}$, the free streaming
length decreases as $t^{-1/3}$.  The maximal streaming length is
$$
\lambda_J^{\rm max} = \lambda^c_J (t_{eq}) \eqno\eq
$$
which for $v (t_{eq}) \sim 0.1$ (appropriate for 25 eV neutrinos)
exceeds the scale of galaxies.
\par
In inflationary Universe models and in the texture theory, the density
perturbations are essentially dark matter fluctuations.  The above free
streaming analysis then shows that, if the dark matter is hot, then
no perturbations on the scale of galaxies
will survive independent of larger-scale structures.  Hence, these
theories are acceptable only if most of the dark matter is cold.
\par
Cosmic string theories, in contrast, work well - if not even better -
with hot dark matter$^{183,184,176)}$.  The cosmic string seeds survive free
streaming.  The growth of perturbations on small scales $\lambda$ is
delayed (it starts once $\lambda = \lambda_J (t)$) but not
prevented.
\par
Let us summarize the main characteristics of the cosmic string, global
texture, and inflationary Universe theories of structure formation.
Inflation predicts random phase perturbations.  The density peaks will
typically be spherical, and the model is consistent with basic
observations only for CDM.  The global texture and cosmic string
models both give non-random phase perturbations.  The topology is
dominated by spherical peaks for textures, whereas it is planar or
filamentary for
cosmic strings (depending on the small scale structure on the
strings).  Textures require CDM, whereas cosmic strings work
better with HDM.

\section{Large-Scale Structure Signatures for Cosmic Strings and
Textures}
\par
The \underbar{genus curve}$^{185)}$ is a statistical measure for the topology
of
large-scale structure.  Given a smooth density field $\rho
(\undertext{x})$, we pick a density $\rho_0$ and consider the surface
$S_{\rho_0}$ where $\rho (\undertext{x}) = \rho_0$ and calculate the
genus $\nu (S)$ of this surface
$$
\nu =  {\rm \# \> of \> holes \> of \>} S - {\rm \# of \> disconnected
\> components \> of \>} S \, . \eqno\eq
$$
The genus curve is the graph of $\nu$ as a function of $\rho_0$.
\par
For perturbations with Poisson statistics, the genus curve can be
calculated analytically (Fig. 31).  The inflationary CDM model in the
linear regime falls in this category.  The genus curve is symmetric
about the mean density $\bar \rho$.  In the texture model, the
symmetry about $\bar \rho$ is broken and the genus curve is shifted to
the left$^{186)}$.  In the cosmic string model, there is a pronounced
asymmetry between $\rho > \bar \rho$ and $\rho < \bar \rho$.  At small
densities, the genus curve measures the (small number) of large voids,
whereas for $\rho > \bar \rho$ the curve picks$^{187)}$ out the large
number of high density peaks which result as a consequence of the
fragmentation of the wakes (Fig. 31).

\midinsert \vskip 10cm
\hsize=6in \raggedbottom
\noindent{\bf Figure 31.} The genus curve of the smoothed mass density
field in a cosmic string wake toy model compared to the symmetric
curve which results in the case of a model with a random distribution
of mass points. The vertical axis is the genus (with genus zero at the
height of the ``x"), the horizontal axis is a measure of density (``0"
denotes average density).
\endinsert

The \undertext{counts in cell statistic}$^{188)}$ can be successfully
applied to distinguish between distributions of galaxies with the same
power spectrum but with different phases.  The statistic is obtained
by dividing the sample volume into equal size cells, counting the
number $f(n)$ of cells containing $n$ galaxies, and plotting $f(n)$ as
a function of $n$.

We$^{189)}$ have applied this statistic to a set of toy models of
large-scale structure.  In each case, the sample volume was (150
Mpc)$^3$, the cell size (3.75 Mpc)$^3$, and the samples contained
90,000 galaxies.  We compared a texture model (galaxies distributed in
spherical clumps separated by 30 Mpc with a Gaussian radial density
field of width 9 Mpc), a cosmic string model dominated by filaments
(all galaxies randomly distributed in filaments of dimensions (60
$\times$ 4 $\times$ 4) Mpc$^3$ with mean separation 30 Mpc, a cosmic
string wake model (same separation and wake dimensions $(40 \times 40
\times 2)$ Mpc$^3$), a cold dark matter model (obtained by Fourier
transforming the CDM power spectrum and assigning random phases), and
a Poisson distribution of galaxies.

\midinsert \vskip 17cm
\hsize=6in \raggedbottom
\noindent{\bf Figure 32.}
The three dimensional counts in cell statistics for a Poisson model
(G), a cold dark matter model (CDM), cosmic string wakes (SW), string
filaments (SF) and textures (T).
\endinsert

As shown in Fig. 32, the predicted curves differ significantly,
demonstrating that this statistic is an excellent one at
distinguishing different theories with the same power spectrum.  The
counts in cell statistic can also be applied to effectively two
dimensional surveys such as single slices of the CFA redshift
survey$^{11)}$.  The predictions of our theoretical toy models are
shown in Fig. 8.

\midinsert \vskip 17cm
\hsize=6in \raggedbottom
\noindent{\bf Figure 33.}
The two dimensional counts in cell statistic for a slice of the
Universe of the dimensions of a CFA slice, evaluated for the same
models as in Fig. 32.
\endinsert

A third statistic which has proved useful$^{190)}$ is distinguishing
models with Gaussian and non-Gaussian phases is the void probability
function $p(R)$, the probability that a sphere of radius $R$ contains
no galaxies.

\section{Signatures in the Microwave Background}

Inflationary Universe models predict essentially random phase
fluctuations in the microwave background with a scale invariant
spectrum $(n = 1)$.  Small deviations from scale invariance are model
dependent and have recently been discussed in detail in Refs. 148.  In all
models, the amplitude must be consistent with
structure formation.  As mentioned in Section 6.1, the COBE
discovery$^{12)}$ of anisotropies in the CMB has provided severe
constraints on inflationary models.  They are only consistent with the
present data if the bias parameter $b$ is about 1, which must be
compared to the value $b \simeq 2.5$ which is the best value for
galaxy formation in this model$^{191)}$.  Note that full sky coverage
is not essential for testing inflationary models since in any set of
local observations of $\delta T/T$, the results will form a Gaussian
distribution about the r.m.s. value.  Mixed dark matter models do
slightly better and have recently been studied vigorously.

\midinsert \vskip 7cm
\hsize=6in \raggedbottom
\noindent{\bf Figure 34.}
Sketch of the mechanism producing linear discontinuities in the
microwave temperature for photons $\gamma$ passing on different sides
of a moving string $S$ (velocity $v$).  $\theta$ is the observer.
Space perpendicular to the string is conical (deficit angle $\alpha$).
\endinsert

Cosmic string models predict non-Gaussian temperature anisotropies.
One mechanism gives rise to localized linear temperature
discontinuities$^{192)}$; its origin is illustrated in Fig. 34.  Photons
passing on different sides of a long straight string moving with
velocity $v$ reach the observer with a Doppler shift
$$
{\delta T\over T} \sim 8 \pi G \mu v \gamma (v) \, . \eqno\eq
$$
To detect such discontinuities, an appropriate survey strategy ({
e.g.}, full sky survey) with small angular resolution is crucial.  The
distribution of strings also gives rise to Sachs-Wolfe type
anisotropies$^{193)}$.

The theoretical error bars in the normalization of CMB anisotropies
from strings are rather large -- a direct consequence of the fact that
the precise form of the scaling solution for the string network is not
well determined.  Nevertheless, we can consider a fixed set of cosmic
string parameters and ask whether the normalizations of $G\mu$ from
large-scale structure data and from COBE are consistent.  This has
been done numerically in Ref. 149, and using an analytical toy model in
Ref. 151.

The analytical model$^{151)}$ is based on adding up as a random walk the
individual Doppler shifts from strings which the microwave photons
separated by angular scale $v$ pass on different sides, and
using this method to compute $\Delta T/T (v)$.  Using the
Bennett-Bouchet$^{194)}$ string parameters, the result for $G \mu$
becomes
$$
G\mu = (1.3 \pm 0.5) 10^{-6} \, ,  \eqno\eq
$$
in good agreement with the requirements from large-scale structure
formation$^{176)}$.

To detect the predicted anisotropies from textures, it is again
essential to have a full sky survey.  However, angular
resolution is adequate this time, since the specific signature for
textures is a small number $( \sim 10)$ of hot and cold disks with
amplitude$^{195)}$
$$
{\delta T\over T} \sim 0.06 \times 16 \> \pi \> G \eta^2 \sim 3 \cdot 10^{-5}
\eqno\eq
$$
and angular size of about $10^\circ$.
The hot and cold spots are due to photons
falling into the expanding Goldstone boson radiation field which
results after texture collapse or due to photons climbing out of the
potential well of the collapsing texture$^{196)}$ (see Fig. 35).

\midinsert \vskip 5cm
\hsize=6in \raggedbottom
\noindent{\bf Figure 35.} Space-time diagram of a collapsing texture
(backward light cone) and the resulting expanding Goldstone boson
radiation (forward light cone). Unwinding of the texture occurs at
point ``TX". The light ray $\gamma_2$ falls into the potential well
and is blueshifted, the ray $\gamma_1$ is redshifted.
\endinsert

Note that the texture model is not ruled out by the recent COBE
results.  The amplitude (6.36) is lower than the pixel sensitivity of
the COBE maps.  However, the predicted quadrupole CMB anisotropy
(normalizing $\eta$ by the large-scale structure data) exceeds the
COBE data point by a factor of between 2.5 and 4$^{150,198)}$.  Hence,
biasing must be invoked in order to try to explain the large-scale
structure data given the reduced value of $\eta$ mandated by the
discovery of CMB anisotropies.

\section{Conclusions}

Topological defect models of structure formation generically give rise
to a scale invariant power spectrum and are hence in good agreement
with the recent COBE results on anisotropies of the CMB.  The
amplitude of the quadrupole temperature fluctuation can be used to
normalize the models.  For cosmic strings, the resulting normalization
agrees well with the normalization from large-scale structure data.
For textures, there is a mismatch which requires introducing biasing.
For textures, the situation is comparable to that in the CDM model,
where COBE demands a bias parameter $b \simeq 1$, whereas galaxy
formation is said to demand$^{191)}$ $b \simeq 2.5$.

It was emphasized that r.m.s. data intrinsically is unable to
differentiate between topological defect models (with non-random
phases) and inflationary modes (with random phases).  We need
statistics which are sensitive to nonrandom phases.

The most economical model for structure formation may be the model
based on cosmic strings and hot dark matter.  It requires no new
particles (although it does require a finite neutrino mass), it agrees
well with COBE and with the CFA redshift data, and it has clear
signatures both for large-scale structure and CMB statistics.
\bigskip
\chapter{Acknowledgments}
\par
I am grateful to Prof. Oscar Eboli for inviting me to give these lectures.
I thank the members of the organizing committee for a lot of help during my
visit to Brazil, and the students at the school for their enthusiasm.

This work has been supported in part by the US Department of Energy under
grant DE FG02-91ER40688, Task A.
\bigskip
\REF\one{A. Linde, `Particle Physics and Inflationary Cosmology'
(Harwood, Chur, 1990).}
\REF\two{S. Blau and A. Guth, `Inflationary Cosmology,' in `300 Years
of Gravitation' ed. by S. Hawking and W. Israel (Cambridge Univ.
Press, Cambridge, 1987).}
\REF\three{K. Olive, {\it Phys. Rep.} {\bf 190}, 307 (1990).}
\REF\four{T.W.G. Kibble, {\it Phys. Rep.} {\bf 67}, 183 (1980).}
\REF\five{A. Vilenkin, {\it Phys. Rep.} {\bf 121}, 263 (1985).}
\REF\six{N. Turok, `Phase Transitions as the Origin of Large-Scale
Structure,' in `Particles, Strings and Supernovae' (TASI-88) ed. by
A. Jevicki and C.-I. Tan (World Scientific, Singapore, 1989).}
\REF\seven{A. Vilenkin and E.P.S. Shellard, `Topological Defects and Cosmology'
(Cambridge Univ. Press, Cambridge, 1993).}
\REF\eight{R. Brandenberger, {\it Rev. Mod. Phys.} {\bf 57}, 1
(1985).}
\REF\nine{R. Brandenberger, in `Physics of the Early Universe,' proc.
of the 1989 Scottish Univ. Summer School in Physics, ed. by J. Peacock, A.
Heavens and A. Davies (SUSSP Publ., Edinburgh, 1990).}
\REF\ten{R. Brandenberger, in proc. of the 1991 Trieste Summer School
in Particle Physics and Cosmology, eds. E. Gava et al. (World
Scientific, Singapore, 1992).}
\REF\eleven{V. de Lapparent, M. Geller and J. Huchra, {\it Ap. J.
(Lett)} {\bf 302}, L1 (1986).}
\REF\twelve{G. Smoot et al., {\it Ap. J. (Lett.)} {\bf 396}, L1
(1992).}
\REF\thirteen{A. Guth, {\it Phys. Rev.} {\bf D23}, 347 (1981).}
\REF\fourteen{A. Linde, {\it Phys. Lett.} {\bf 129B}, 177 (1983).}
\REF\fifteen{V. Mukhanov, H. Feldman and R. Brandenberger, {\it Phys.
Rep.} {\bf 215}, 203 (1992).}
\REF\sixteen{ T.W.B. Kibble, {\it J. Phys.} {\bf A9}, 1387 (1976).}
\REF\seventeen{V. Mukhanov and R. Brandenberger, {\it Phys. Rev.
Lett.} {\bf 68}, 1969 (1992).}
\REF\eightteen{R. Brandenberger, `Towards a Nonsingular Universe,' in
Proc. of the TEXAS-PASCOS meeting, Berkeley, Dec. 12-18, 1992 (Ann. N.Y.
Academy of Sciences, 1993)}
\REF\nineteen{E. Milne, {\it Zeits. f. Astrophys.} {\bf 6}, 1 (1933).}
\REF\twenty{S. Shechtman, P. Schechter, A. Oemler, D. Tucker, R. Kirshner and
H. Lin, Harvard-Smithsonian preprint CFA 3385 (1992), to appear in `Clusters
and Superclusters of Galaxies', ed. by A. Fabian (Kluwer, Dordrecht, 1993).}
\REF\twentyone{see e.g., S. Weinberg, `Gravitation and Cosmology'
(Wiley, New York, 1972); \nextline
Ya.B. Zel'dovich and I. Novikov, `The Structure and Evolution of the
Universe' (Univ. of Chicago Press, Chicago, 1983).}
\REF\twentytwo{E. Hubble, {\it Proc. Nat. Acad. Sci.} {\bf 15}, 168
(1927).}
\REF\twentythree{R. Alpher and R. Herman, {\it Rev. Mod. Phys.} {\bf
22}, 153 (1950); \nextline
G. Gamov, {\it Phys. Rev.} {\bf 70}, 572 (1946).}
\REF\twentyfour{R. Alpher, H. Bethe and G. Gamov, {\it Phys. Rev.}
{\bf 73}, 803 (1948); \nextline
R. Alpher and R. Herman, {\it Nature} {\bf 162}, 774 (1948).}
\REF\twentyfive{For an excellent introduction see S. Weinberg, `The
First Three Minutes' (Basic Books, New York, 1988).}
\REF\twentysix{R. Dicke, P.J.E. Peebles, P. Roll and D. Wilkinson,
{\it Ap. J.} {\bf 142}, 414 (1965).}
\REF\twentyseven{A. Penzias and R. Wilson, {\it Ap. J.} {\bf 142}, 419
(1965).}
\REF\twentyeight{J. Mather et al., {\it Ap. J. (Lett.)} {\bf 354}, L37
(1990).}
\REF\twentynine{H. Gush, M. Halpern and E. Wishnow, {\it Phys. Rev.
Lett.} {\bf 65}, 937 (1990).}
\REF\thirty{see e.g., G. Efstathiou, in `Physics of the Early
Universe,' proc. of the 1989 Scottish Univ. Summer School in Physics,
ed. by J. Peacock, A. Heavens and A. Davies (SUSSP Publ., Edinburgh,
1990).}
\REF\thirtyone{B. Pagel, {\it Ann. N.Y. Acad. Sci.} {\bf 647}, 131
(1991).}
\REF\thirtytwo{H. Arp, G. Burbidge, F. Hoyle, J. Narlikar and N.
Vickramasinghe, {\it Nature} {\bf 346}, 807 (1990).}
\REF\thirtythree{N. Tsamis and R. Woodard, `Relaxing the Cosmological
Constant', {\it Phys. Lett. B}, in press (1993).}
\REF\thirtyfour{J. Tonry, talk at the TEXAS-PASCOS meeting, Berkeley,
Dec. 12-18, 1992, to be publ. in the proceedings (Ann. N.Y. Academy of
Sciences, 1993).}
\REF\thirtyfive{V. Trimble, {\it Ann. Rev. Astr. Astrophys.} {\bf 25}, 423
(1987).}
\REF\thirtysix{E. Bertschinger and A. Dekel, {\it Ap. J.} {\bf 336},
L5 (1989).}
\REF\thirtyseven{M. Strauss, M. Davis, A. Yahil and J. Huchra, {\it
Ap. J.} {\bf 385}, 421 (1992).}
\REF\thirtyeight{G. Ellis, {\it Class. Quant. Grav.} {\bf 5}, 207 (1988);
\nextline
G. Ellis, D. Lyth and M. Mijic, {\it Phys. Lett.} {\bf 271B}, 52
(1991);\nextline
M. Madsen, G. Ellis, J. Mimoso and J. Butcher, {\it Phys. Rev.} {\bf D46}, 1399
(1992).}
\REF\thirtynine{J. Primack, D. Seckel and B. Sadoulet, {\it Ann. Rev. Nucl.
Part. Sci.} {\bf 38}, 751 (1988).}
\REF\forty{L. da Costa, in `The Distribution of Matter in the
Universe' ed. by D. Gerbal and G. Mamon, in press (1991).}
\REF\fortyone{T. Broadhurst, R. Ellis, D. Koo and A. Szalay, {\it
Nature} {\bf 343}, 726 (1990).}
\REF\fortytwo{J. Ostriker and L. Cowie, {\it Ap. J. (Lett.)} {\bf
243}, L127 (1981).}
\REF\fortythree{D. Kazanas, {\it Ap. J.} {\bf 241}, L59 (1980).}
\REF\fortyfour{W. Press, {\it Phys. Scr.} {\bf 21}, 702 (1980).}
\REF\fortyfive{G. Chibisov and V. Mukhanov, `Galaxy Formation and
Phonons,' Lebedev Physical Institute Preprint No. 162 (1980);
\nextline
G. Chibisov and V. Mukhanov, {\it Mon. Not. R. Astron. Soc.} {\bf
200}, 535 (1982).}
\REF\fortysix{V. Lukash, {\it Pis'ma Zh. Eksp. Teor. Fiz.} {\bf 31}, 631
(1980).}
\REF\fortyseven{K. Sato, {\it Mon. Not. R. Astron. Soc.} {\bf 195},
467 (1981).}
\REF\fortyeight{D. Kirzhnits and A. Linde, {\it Pis'ma Zh. Eksp.
Teor. Fiz.} {\bf 15}, 745 (1972); \nextline
D. Kirzhnits and A. Linde, {\it Zh. Eksp. Teor. Fiz.} {\bf 67}, 1263
(1974);\nextline
C. Bernard, {\it Phys. Rev.} {\bf D9}, 3313 (1974);\nextline
L. Dolan and R. Jackiw, {\it Phys. Rev.} {\bf D9}, 3320 (1974);\nextline
S. Weinberg, {\it Phys. Rev.} {\bf D9}, 3357 (1974).}
\REF\fortynine{G. Mazenko, W. Unruh and R. Wald, {\it Phys. Rev.} {\bf
D31}, 273 (1985).}
\REF\fifty{A. Linde, {\it Phys. Lett.} {\bf 108B}, 389 (1982);
\nextline
A. Albrecht and P. Steinhardt, {\it Phys. Rev. Lett.} {\bf 48}, 1220
(1982).}
\REF\fone{J. Langer, {\it Physica} {\bf 73}, 61 (1974).}
\REF\ftwo{S. Coleman, {\it Phys. Rev.} {\bf D15}, 2929 (1977);
\nextline
C. Callan and S. Coleman, {\it Phys. Rev.} {\bf D16}, 1762 (1977).}
\REF\fthree{M. Voloshin, Yu. Kobzarev and L. Okun, {\it Sov. J. Nucl.
Phys.} {\bf 20}, 644 (1975).}
\Ref\one{M. Stone, {\it Phys. Rev.} {\bf D14}, 3568 (1976);\nextline
M. Stone, {\it Phys. Lett.} {\bf 67B}, 186 (1977).}
\Ref\one{P. Frampton, {\it Phys. Rev. Lett.}, {\bf 37}, 1380 (1976).}
\Ref\one{S. Coleman, in `The Whys of Subnuclear Physics' (Erice 1977), ed by
A. Zichichi (Plenum, New York, 1979).}
\REF\fiftyseven{A. Guth and S.-H. Tye, {\it Phys. Rev. Lett.} {\bf
44}, 631 (1980).}
\Ref\one{A. Guth and E. Weinberg, {\it Nucl. Phys.} {\bf B212}, 321 (1983).}
\Ref\one{S. Hawking and I. Moss, {\it Phys. Lett.} {\bf 110B}, 35 (1982).}
\Ref\one{R. Matzner, in `Proceedings of the Drexel Workshop on Numerical
Relativity', ed by J. Centrella (Cambridge Univ. Press, Cambridge, 1986).}
\Ref\one{A. Albrecht, P. Steinhardt, M. Turner and F. Wilczek, {\it Phys. Rev.
Lett.} {\bf 48}, 1437 (1982).}
\Ref\one{L. Abbott, E. Farhi and M. Wise, {\it Phys. Lett.} {\bf 117B}, 29
(1982).}
\Ref\one{J. Traschen and R. Brandenberger, {\it Phys. Rev.} {\bf D42},
2491 (1990).}
\REF\one{L. Landau and E. Lifshitz, `Mechanics' (Pergamon, Oxford,
1960); \nextline
V. Arnold, `Mathematical Methods of Classical Mechanics' (Springer,
New York, 1978).}
\REF\one{S. Coleman and E. Weinberg, {\it Phys. Rev.} {\bf D7}, 1888
(1973).}
\REF\one{M. Markov and V. Mukhanov, {\it Phys. Lett.} {\bf 104A}, 200
(1984); \nextline
V. Belinsky, L. Grishchuk, I. Khalatnikov and Ya. Zel'dovich, {\it
Phys. Lett.} {\bf 155B}, 232 (1985); \nextline
L. Kofman, A. Linde and A. Starobinsky, {\it Phys. Lett.} {\bf 157B},
36 (1985); \nextline
T. Piran and R. Williams, {\it Phys. Lett.} {\bf 163B}, 331 (1985).}
\REF\one{S.-Y. Pi, {\it Phys. Rev. Lett.} {\bf 52}, 1725 (1984).}
\REF\one{D. Nanopoulos, K. Olive, M. Srednicki and K. Tamvakis, {\it
Phys. Lett.} {\bf 123B}, 41 (1983); \nextline
J. Ellis, K. Enqvist, D. Nanopoulos, K. Olive and M. Srednicki, {\it
Phys. Lett.} {\bf 152B}, 175 (1985); \nextline
R. Holman, P. Ramond and C. Ross, {\it Phys. Lett.} {\bf 137B}, 343
(1984).}
\REF\one{A. Goncharov and A. Linde, {\it JETP} {\bf 59}, 930 (1984);
\nextline
A. Goncharov and A. Linde, {\it Phys. Lett.} {\bf 139B}, 27 (1984);
\nextline
A. Goncharov and A. Linde, {\it Class. Quant. Grav.} {\bf 1}, L75
(1984).}
\REF\one{I. Antoniadis, J. Ellis, J. Hagelin and D. Nanopoulos, {\it
Phys. Lett.} {\bf 205B}, 459 (1988); \nextline
I. Antoniadis, J. Ellis, J. Hagelin and D. Nanopoulos, {\it Phys.
Lett.} {\bf 208B}, 209 (1988).}
\REF\one{A. Linde, {\it Phys. Lett.} {\bf 162B}, 281 (1985).}
\REF\one{A. Starobinsky, {\it Phys. Lett.} {\bf 91B}, 99 (1980).}
\REF\one{M. Mijic, M. Morris and W.-M. Suen, {\it Phys. Rev.} {\bf
D34}, 2934 (1986).}
\REF\one{B. Whitt, {\it Phys. Lett.} {\bf 145B}, 176 (1984).}
\REF\one{C. Mathiozhagan and V. Johri, {\it Class. Quant. Grav.} {\bf
1}, L24 (1984).}
\REF\one{D. La and P. Steinhardt, {\it Phys. Rev. Lett.} {\bf 62}, 376
(1989).}
\REF\one{E. Weinberg, {\it Phys. Rev.} {\bf D40}, 3950 (1989).}
\REF\one{D. La, P. Steinhardt and E. Bertschinger, {\it Phys. Lett.}
{\bf B231}, 231 (1989).}
\REF\one{F. Adams, K. Freese and A. Guth, {\it Phys. Rev.} {\bf D43},
965 (1991).}
\REF\one{H. Feldman and R. Brandenberger, {\it Phys. Lett.} {\bf
227B}, 359 (1989).}
\REF\one{J. Kung and R. Brandenberger, {\it Phys. Rev.} {\bf D42},
1008 (1990).}
\REF\one{D. Goldwirth and T. Piran, {\it Phys. Rev. Lett.} {\bf 64},
2852 (1990);\nextline
D. Goldwirth and T. Piran, {\it Phys. Rep.} {\bf 214}, 223 (1992).}
\REF\one{A. Albrecht and R. Brandenberger, {\it Phys. Rev.} {\bf D31},
1225 (1985).}
\REF\one{R. Brandenberger, R. Laflamme and M. Mijic, {\it Mod. Phys.
Lett.} {\bf A5}, 2311 (1990).}
\REF\one{R. Brandenberger, H. Feldman, V. Mukhanov and T. Prokopec,
`Gauge Invariant Cosmological Perturbations: Theory and Applications,'
Brown preprint BROWN-HET-860 (1992), to be publ. in ``The Origins of
Structure in the Universe," Pont d'Oye Workshop Proceedings, April 26-
May 2, 1992.}
\REF\one{J. Stewart, {\it Class. Quantum Grav.} {\bf 7}, 1169 (1990).}
\REF\one{J. Stewart and M. Walker, {\it Proc. R. Soc.} {\bf A341}, 49
(1974).}
\REF\one{J. Bardeen, {\it Phys. Rev.} {\bf D22}, 1882 (1980).}
\REF\one{R. Brandenberger, R. Kahn and W. Press, {\it Phys. Rev.} {\bf
D28}, 1809 (1983).}
\REF\one{H. Kodama and M. Sasaki, {\it Prog. Theor. Phys. Suppl.} No.
78, 1 (1984).}
\REF\one{R. Durrer and N. Straumann, {\it Helvet. Phys. Acta} {\bf
61}, 1027 (1988).}
\REF\one{D. Lyth and M. Mukherjee, {\it Phys. Rev.} {\bf D38}, 485
(1988).}
\REF\one{G.F.R. Ellis and M. Bruni, {\it Phys. Rev.} {\bf D40}, 1804
(1989).}
\REF\one{G. Chibisov and V. Mukhanov, {\it JETP Lett.} {\bf 33}, 532
(1981);\nextline
V. Mukhanov and G. Chibisov, {\it Zh. Eksp. Teor. Fiz.} {\bf 83}, 475
(1982).}
\REF\one{A. Lapedes, {\it J. Math. Phys.} {\bf 19}, 2289 (1978);
\nextline
R. Brandenberger and R. Kahn, {\it Phys. Lett.} {\bf 119B}, 75
(1982).}
\REF\one{J. Bardeen, unpublished (1984).}
\REF\one{R. Brandenberger, {\it Nucl. Phys.} {\bf B245}, 328 (1984).}
\REF\one{L. Landau and E. Lifshitz, `Theoretical Physics, Vol. II:
Classical Fields' (Pergamon, London, 1958).}
\REF\one{N. Birrell and P. Davies, `Quantum Fields in Curved Space'
(Cambridge Univ. Press, Cambridge, 1982).}
\REF\one{R. Brandenberger and C. Hill, {\it Phys. Lett.} {\bf 179B},
30 (1986).}
\REF\one{A. Vilenkin and L. Ford, {\it Phys. Rev. } {\bf D26}, 1231
(1982); \nextline
A. Linde, {\it Phys. Lett.} {\bf 116B}, 335 (1982).}
\REF\one{R. Kahn and R. Brandenberger, {\it Phys. Lett.} {\bf 141B},
317 (1984).}
\REF\one{A. Guth and S.-Y. Pi, {\it Phys. Rev. Lett.} {\bf 49}, 1110
(1982); \nextline
S. Hawking, {\it Phys. Lett.} {\bf 115B}, 295 (1982); \nextline
A. Starobinsky, {\it Phys. Lett.} {\bf 117B}, 175 (1982); \nextline
J. Bardeen, P. Steinhardt and M. Turner, {\it Phys. Rev.} {\bf D28},
679 (1983); \nextline
R. Brandenberger and R. Kahn, {\it Phys. Rev.} {\bf D28}, 2172 (1984);
\nextline
J. Frieman and M. Turner, {\it Phys. Rev.} {\bf D30}, 265 (1984);
\nextline
V. Mukhanov, {\it JETP Lett.} {\bf 41}, 493 (1985).}
\REF\one{ `Quantum Theory and Measurement,' ed. J. Wheeler and W.
Zurek (Princeton Univ. Press, Princeton, 1983).}
\REF\one{W. Zurek, {\it Phys. Rev.} {\bf D24}, 1516 (1982); \nextline
W. Zurek, {\it Phys. Rev.} {\bf D26} 1862 (1982).}
\REF\one{E. Joos and H. Zeh, {\it Z. Phys.} {\bf B59}, 223 (1985);
\nextline
H. Zeh, {\it Phys. Lett.} {\bf 116A}, 9 (1986); \nextline
C. Kiefer, {\it Class. Quantum Grav.} {\bf 4}, 1369 (1987); \nextline
T. Fukuyama and M. Morikawa, {\it Phys. Rev.} {\bf D39}, 462 (1989);
\nextline
J. Halliwell, {\it Phys. Rev.} {\bf D39}, 2912 (1989); \nextline
T. Padmanabhan, {\it Phys. Rev.} {\bf D39}, 2924 (1980); \nextline
W. Unruh and W. Zurek, {\it Phys. Rev.} {\bf D40}, 1071 (1989);
\nextline
E. Calzetta and F. Mazzitelli, {\it Phys. Rev.} {\bf D42}, 4066
(1990); \nextline
S. Habib and R. Laflamme, {\it Phys. Rev.} {\bf D42}, 4056 (1990);
\nextline
H. Feldman and A. Kamenshchik, {\it Class. Quantum Grav.} {\bf 8}, L65
(1991).}
\REF\one{M. Sakagami, {\it Prog. Theor. Phys.} {\bf 79}, 443 (1988).}
\REF\one{E. Harrison, {\it Phys. Rev.} {\bf D1}, 2726 (1970);
\nextline
Ya.B. Zel'dovich, {\it Mon. Not. R. astron. Soc.} {\bf 160}, 1p
(1972).}
\REF\one{W. Fischler, B. Ratra and L. Susskind, {\it Nucl. Phys.} {\bf
B259}, 730 (1985).}
\REF\one{see e.g., G. Ross, Grand Unified Theories (Benjamin, Reading,
1985).}
\REF\one{Ya.B. Zel'dovich, I. Kobzarev and L. Okun, {\it Zh. Eksp.
Teor. Fiz.} {\bf 67}, 3 (1974).}
\REF\one{Ya.B. Zel'dovich and M. Khlopov, {\it Phys. Lett.} {\bf 79B},
239 (1978); \nextline
J. Preskill, {\it Phys. Rev. Lett.} {\bf 43}, 1365 (1979).}
\REF\one{P. Langacker and S.-Y. Pi, {\it Phys. Rev. Lett.} {\bf 45}, 1
(1980).}
\REF\one{T.W.B. Kibble and E. Weinberg, {\it Phys. Rev.} {\bf D43},
3188 (1991).}
\REF\one{M. Barriola and A. Vilenkin, {\it Phys. Rev. Lett.} {\bf 63},
341 (1989).}
\REF\one{S. Rhie and D. Bennett, {\it Phys. Rev. Lett.} {\bf 65}, 1709
(1990).}
\REF\one{H. Nielsen and P. Olesen, {\it Nucl. Phys.} {\bf B61}, 45
(1973).}
\REF\one{R. Davis, {\it Phys. Rev.} {\bf D35}, 3705 (1987).}
\REF\one{N. Turok, {\it Phys. Rev. Lett.} {\bf 63}, 2625 (1989).}
\REF\one{T.W.B. Kibble, {\it Acta Physica Polonica} {\bf B13}, 723
(1982).}
\REF\one{T. Vachaspati and A. Vilenkin, {\it Phys. Rev.} {\bf D30},
2036 (1984).}
\REF\one{S. Rudaz and A. Srivastava, `On the Production of Flux
Vortices and Magnetic Monopoles in Phase Transitions,' Univ. of
Minnesota preprint UMN-TH-1028/92 (1992).}
\REF\one{J. Ye and R. Brandenberger, {\it Nucl. Phys.} {\bf B346}, 149
(1990).}
\REF\one{M. Hindmarsh, A.-C. Davis and R. Brandenberger, `Formation of
Topological Defects in First and Second Order Phase Transitions,' Brown Univ.
preprint BROWN-HET-902 (1993).}
\REF\one{Ya.B. Zel'dovich, {\it Mon. Not. R. astron. Soc.} {\bf 192},
663 (1980).}
\REF\one{A. Vilenkin, {\it Phys. Rev. Lett.} {\bf 46}, 1169 (1981).}
\REF\one{T. Prokopec, A. Sornborger and R. Brandenberger, {\it Phys.
Rev.} {\bf D45}, 1971 (1992).}
\REF\one{J. Borrill, E. Copeland and A. Liddle, {\it Phys. Lett.} {\bf 258B},
310 (1991).}
\REF\one{A. Sornborger, `A Semi-Analytic Study of Texture
Collapse,' Brown Univ. preprint BROWN-HET-895 (1993).}
\REF\one{L. Perivolaropoulos, {\it Phys. Rev.} {\bf D46}, 1858
(1992).}
\REF\one{T. Prokopec, {\it Phys. Lett.} {\bf 262B}, 215 (1991);\nextline
R. Leese and T. Prokopec, {\it Phys. Rev.} {\bf D44}, 3749
(1991).}
\REF\one{D. Foerster, {\it Nucl. Phys.} {\bf B81}, 84 (1974).}
\REF\one{N. Turok, in `Proceedings of the 1987 CERN/ESO Winter School
on Cosmology and Particle Physics' (World Scientific, Singapore,
1988).}
\REF\one{T.W.B. Kibble and N. Turok, {\it Phys. Lett.} {\bf 116B}, 141
(1982).}
\REF\one{R. Brandenberger, {\it Nucl. Phys.} {\bf B293}, 812 (1987).}
\REF\one{E.P.S. Shellard, {\it Nucl. Phys.} {\bf B283}, 624 (1987).}
\REF\one{R. Matzner, {\it Computers in Physics} {\bf 1}, 51 (1988);
\nextline
K. Moriarty, E. Myers and C. Rebbi, {\it Phys. Lett.} {\bf 207B}, 411
(1988); \nextline
E.P.S. Shellard and P. Ruback, {\it Phys. Lett.} {\bf 209B}, 262
(1988).}
\REF\one{P. Ruback, {\it Nucl. Phys.} {\bf B296}, 669 (1988).}
\REF\one{T. Vachaspati and A. Vilenkin, {\it Phys. Rev.} {\bf D31},
3052 (1985); \nextline
N. Turok, {\it Nucl. Phys.} {\bf B242}, 520 (1984); \nextline
C. Burden, {\it Phys. Lett.} {\bf 164B}, 277 (1985).}
\REF\one{A. Albrecht and N. Turok, {\it Phys. Rev. Lett.} {\bf 54},
1868 (1985).}
\REF\one{D. Bennett and F. Bouchet, {\it Phys. Rev. Lett.} {\bf 60},
257 (1988).}
\REF\one{B. Allen and E.P.S. Shellard, {\it Phys. Rev. Lett.} {\bf
64}, 119 (1990).}
\REF\one{A. Albrecht and N. Turok, {\it Phys. Rev. } {\bf D40}, 973
(1989).}
\REF\one{R. Brandenberger and J. Kung, in `The Formation and Evolution
of Cosmic Strings' eds. G. Gibbons, S. Hawking and T. Vachaspati
(Cambridge Univ. Press, Cambridge, 1990).}
\REF\one{see e.g., C. Misner, K. Thorne and J. Wheeler, `Gravitation'
(Freeman, San Francisco, 1973).}
\REF\one{B. Carter, {\it Phys. Rev.} {\bf D41}, 3869 (1990).}
\REF\one{E. Copeland, T.W.B. Kibble and D. Austin, {\it Phys. Rev.}
{\bf D45}, 1000 (1992).}
\REF\onefortyeight{L. Krauss and M. White, {\it Phys. Rev. Lett.} {\bf 69}, 869
(1992);\nextline
D. Salopek, {\it Phys. Rev. Lett.} {\bf 69}, 3602 (1992);\nextline
R. Davis, H. Hodges, G. Smoot, P. Steinhardt and M. Turner, {\it Phys. Rev.
Lett.} {\bf 69}, 1856 (1992);\nextline
A. Liddle and D. Lyth, {\it Phys. Lett.} {\bf 291B}, 391 (1992); \nextline
F. Lucchin, S. Matarrese and S. Mollerach, {\it Ap. J. (Lett.)}, {\bf 401}, L49
(1992);\nextline
T. Souradeep and V. Sahni, {\it Mod. Phys. Lett.} {\bf A7}, 3541 (1992).}
\REF\one{D. Bennett, A. Stebbins and F. Bouchet, {\it Ap. J. (Lett.)}
{\bf 399}, L5 (1992).}
\REF\one{D. Bennett and S. Rhie, `COBE's Constraints on the Global Monopole and
Texture Theories of Cosmic Structure Formation', {\it Ap. J. (Lett.)}, in press
(1993).}
\REF\one{L. Perivolaropoulos, {\it Phys. Lett.} {\bf 298B}, 305 (1993).}
\REF\one{R. Sachs and A. Wolfe, {\it Ap. J.} {\bf 147}, 73 (1967).}
\REF\one{A. Lasenby, lecture at the International School `D. Chalonge', 2nd
Course, 6 - 13 Sept. 1992, Erice, Italy}
\REF\one{P. Coles and J. Barrow, {\it Mon. Not. R. astron. Soc.} {\bf 228}, 407
(1987);\nextline
P. Coles, {\it Mon. Not. R. astron. Soc.} {\bf 234}, 501 (1988);\nextline
J. Gott et al., {\it Ap. J.} {\bf 340}, 625 (1989).}
\REF\one{L. Perivolaropoulos, `On the Statistics of CMB Fluctuations induced by
Topological Defects', Harvard-Smithsonian Center for Astrophysics preprint
CFA 3526 (1992); \nextline
R. Moessner, L. Perivolaropoulos and R. Brandenberger, `Statistics of the
Spatial Gradient Map of CMB Fluctuations induced by Topological Defects', Brown
preprint in preparation (1993).}
\REF\one{Ya.B. Zel'dovich, J. Einasto and S. Shandarin, {\it Nature}
{\bf 300}, 407 (1982); \nextline
J. Oort, {\it Ann. Rev. Astron. Astrophys.} {\bf 21}, 373 (1983);
\nextline
R.B. Tully, {\it Ap. J.} {\bf 257}, 389 (1982); \nextline
S. Gregory, L. Thomson and W. Tifft, {\it Ap. J.} {\bf 243}, 411
(1980).}
\REF\one{G. Chincarini and H. Rood, {\it Nature} {\bf 257}, 294
(1975); \nextline
J. Einasto, M. Joeveer and E. Saar, {\it Mon. Not. R. astron. Soc.}
{\bf 193}, 353 (1980); \nextline
R. Giovanelli and M. Haynes, {\it Astron. J.} {\bf 87}, 1355 (1982);
\nextline
D. Batuski and J. Burns, {\it Ap. J.} {\bf 299}, 5 (1985).}
\REF\one{R. Kirshner, A. Oemler, P. Schechter and S. Shechtman, {\it
Ap. J. (Lett.)} {\bf 248}, L57 (1981).}
\REF\one{M. Joeveer, J. Einasto and E. Tago, {\it Mon. Not. R. astron.
Soc.} {\bf 185}, 357 (1978); \nextline
L. da Costa et al., {\it Ap. J.} {\bf 327}, 544 (1988).}
\REF\one{A. Dressler, D. Lynden-Bell, D. Burstein, R. Davies, S.
Faber, R. Terlevich and G. Wegner, {\it Ap. J.} {\bf 313}, 42
(1987);\nextline
C. Collins, R. Joseph and N. Robertson, {\it Nature} {\bf 320}, 506
(1986).}
\REF\one{E. Bertschinger and A. Dekel, {\it Ap. J. (Lett.)} {\bf 336}, L5
(1989).}
\REF\one{M. Aaronson, G. Bothun, J. Mould, R. Schommer and M. Cornell,
{\it Ap. J.} {\bf 302}, 536 (1986); \nextline
J. Lucey and D. Carter, {\it Mon. Not. R. astron. Soc.} {\bf 235},
1177 (1988).}
\REF\one{G. Abell, {\it Ap. J. Suppl.} {\bf 3}, 211 (1958).}
\REF\one{M. Davis and J. Huchra, {\it Ap. J.} {\bf 254}, 437 (1982).}
\REF\one{P.J.E. Peebles, in `Physical Cosmology,' 1979 Les Houches
Lectures, ed. by R. Balian, J. Audouze and D. Schramm (North-Holland,
Amsterdam, 1980).}
\REF\one{N. Bahcall and R. Soneira, {\it Ap. J.} {\bf 270}, 20 (1983);
\nextline
A. Klypin and A. Kopylov, {\it Sov. Astr. Lett.} {\bf 9}, 41 (1983).}
\REF\one{A. Dekel, G. Blumenthal, J. Primack and S. Olivier, {\it Ap.
J. (Lett.)} {\bf 338}, L5 (1989); \nextline
W. Sutherland, {\it Mon. Not. R. astron. Soc.} {\bf 234}, 159 (1988).}
\REF\one{N. Bahcall, {\it Ann. Rev. Astr. \& Astrophys.} {\bf 15}, 505
(1977); \nextline
B. Binggeli, in `Nearly Normal Galaxies,' ed. S. Faber (Springer, New
York, 1986).}
\REF\one{see e.g., E. Athanassoula and A. Bosma, in `Large Scale
Structures of the Universe,' IAU Symposium No. 130, ed. by J. Audouze
et al. (Kluwer, Dordrecht, 1988).}
\REF\one{see e.g., S.M. Fall, in `Internal Kinetics and Dynmaics of
Galaxies,' ed. by E. Athanassoula (Reidel, Dordrecht, 1983).}
\REF\one{N. Turok and R. Brandenberger, {\it Phys. Rev.} {\bf D33},
2175 (1986); \nextline
A. Stebbins, {\it Ap. J. (Lett.)} {\bf 303}, L21 (1986); \nextline
H. Sato, {\it Prog. Theor. Phys.} {\bf 75}, 1342 (1986).}
\REF\one{A. Vilenkin, {\it Phys. Rev.} {\bf D23}, 852 (1981);
\nextline
J. Gott, {\it Ap. J.} {\bf 288}, 422 (1985); \nextline
W. Hiscock, {\it Phys. Rev.} {\bf D31}, 3288 (1985); \nextline
B. Linet, {\it Gen. Rel. Grav.} {\bf 17}, 1109 (1985); \nextline
D. Garfinkle, {\it Phys. Rev.} {\bf D32}, 1323 (1985); \nextline
R. Gregory, {\it Phys. Rev. Lett.} {\bf 59}, 740 (1987).}
\REF\one{J. Silk and V. Vilenkin, {\it Phys. Rev. Lett.} {\bf 53},
1700 (1984).}
\REF\one{T. Vachaspati, {\it Phys. Rev. Lett.} {\bf 57}, 1655 (1986).}
\REF\one{A. Stebbins, S. Veeraraghavan, R. Brandenberger, J. Silk and
N. Turok, {\it Ap. J.} {\bf 322}, 1 (1987).}
\REF\one{R. Brandenberger, L. Perivolaropoulos and A. Stebbins, {\it
Int. J. of Mod. Phys.} {\bf A5}, 1633 (1990); \nextline
L. Perivolarapoulos, R. Brandenberger and A. Stebbins, {\it Phys.
Rev.} {\bf D41}, 1764 (1990); \nextline
R. Brandenberger, {\it Phys. Scripta} {\bf T36}, 114 (1991).}
\REF\one{D. Vollick, {\it Phys. Rev.} {\bf D45}, 1884 (1992);
\nextline
T. Vachaspati and A. Vilenkin, {\it Phys. Rev. Lett.} {\bf 67}, 1057 (1991).}
\REF\one{D. Spergel, N. Turok, W. Press and B. Ryden, {\it Phys. Rev.}
{\bf D43}, 1038 (1991).}
\REF\one{A. Gooding, D. Spergel and N. Turok, {\it Ap. J. (Lett.)}
{\bf 372}, L5 (1991); \nextline
C. Park, D. Spergel and N. Turok, {\it Ap. J. (Lett.)} {\bf 373}, L53 (1991).}
\REF\one{R. Cen, J. Ostriker, D. Spergel and N. Turok, {\it Ap. J.}
{\bf 383}, 1 (1991).}
\REF\one{L. Grishchuk and Y. Sidorov, {\it Class. Quant. Grav.} {\bf
6}, L161 (1989); \nextline
L. Grishchuk and Y. Sidorov, {\it Phys. Rev.} {\bf D42}, 3413 (1990);
\nextline
L. Grishchuk, `Quantum Mechanics of the Primordial Cosmological
Perturbations,' to appear in the proceedings of the Sixth Marcel
Grossmann Meeting (Kyoto, 1991).}
\REF\one{T. Prokopec, `Entropy of the Squeezed Vacuum,' Brown Univ.
preprint BROWN-HET-861 (1992); \nextline
A. Albrecht, P. Ferreira, M. Joyce and T. Prokopec, `Inflation and
Squeezed Quantum States,' Imperial College preprint (1993).}
\REF\one{R. Brandenberger, N. Kaiser, D. Schramm and N. Turok, {\it
Phys. Rev. Lett.} {\bf 59}, 2371 (1987).}
\REF\one{R. Brandenberger, N. Kaiser and N. Turok, {\it Phys. Rev.}
{\bf D36}, 2242 (1987).}
\REF\one{J. Gott, A. Melott and M. Dickinson, {\it Ap. J.} {\bf 306},
341 (1986).}
\REF\one{A. Gooding, C. Park, D. Spergel, N. Turok and J. Gott, {\it
Ap. J.} {\bf 393}, 42 (1992).}
\REF\one{J. Gerber and R. Brandenberger, `Topology of Large-Scale
Structure in a Cosmic String Wake Model,' Brown Univ. preprint
BROWN-HET-829 (1991).}
\REF\one{W. Saslaw, {\it Ap. J.} {\bf 297}, 49 (1985).}
\REF\one{S. Ramsey, Senior thesis, Brown Univ. (1992); \nextline
S. Ramsey and R. Brandenberger, in preparation (1993).}
\REF\one{E. Valentini and R. Brandenberger, in preparation (1993);
\nextline
D. Weinberg and S. Cole, `Non-Gaussian Fluctuations and the Statistics
of Galaxy Clustering,' Berkeley preprint CfPA-TH-91-025 (1991).}
\REF\one{S. White, C. Frenk, M. Davis and G. Efstathiou, {\it Ap. J.}
{\bf 313}, 505 (1987).}
\REF\one{N. Kaiser and A. Stebbins, {\it Nature} {\bf 310}, 391
(1984).}
\REF\one{J. Traschen, N. Turok and R. Brandenberger, {\it Phys. Rev.}
{\bf D34}, 919 (1986); \nextline
S. Veeraraghavan and A. Stebbins, `Large-Scale Microwave Anisotropy
from Gravity Seeds,' Fermilab preprint 92/147-A (1992).}
\REF\one{D. Bennett and F. Bouchet, {\it Phys. Rev.} {\bf D41}, 2408
(1990).}
\REF\one{R. Durrer and D. Spergel, `Microwave Anisotropies from
Texture Seeded Structure Formation,' Princeton preprint PUPT-91-1247
(1991).}
\REF\oneninetyseven{N. Turok and D. Spergel, {\it Phys. Rev. Lett.} {\bf 64},
2736 (1990).}
\REF\one{U.-L. Pen, D. Spergel and N. Turok, `Cosmic Structure Formation and
Microwave Anisotropies from Global Field Ordering, Princeton preprint
PU-TH-1375 (1993).}
\refout